\documentclass[12pt]{article}

\usepackage{a4}
\usepackage[reqno,sumlimits]{amsmath}
\usepackage{amsfonts,amssymb} 
\usepackage{theorem} 
\renewcommand{\Im}{\mathrm{Im}\,}

\DeclareMathOperator{\L2}{L^{2}}
\DeclareMathOperator{\supp}{supp}
\DeclareMathOperator{\Tr}{Tr}

\DeclareMathOperator{\one}{\mathbf{1}}

\def\bbC{\mathbb{C}}
\def\bbE{\mathbb{E}}
\def\bbN{\mathbb{N}}
\def\bbP{\mathbb{P}}
\def\bbR{\mathbb{R}}
\def\bbZ{\mathbb{Z}}
\def\d{\rm d}

\def\calC{\mathcal{C}}
\def\calD{\mathcal{D}}
\def\calH{\mathcal{H}}

\def\bE{\rm {\bf {E}}}
\newcommand{\gF}{{\mathfrak F}}
\newcommand{\gH}{{\mathfrak H}}
\newcommand{\gW}{{\mathfrak W}}

\def\supp{{\rm supp}}
\def\dist{{\rm dist}}
\def\unc{{\mathbf{1}}}
\newcommand{\eps}{\varepsilon}
\newcommand{\vp}{\varphi}
\newcommand{\la}{\langle}
\newcommand{\ra}{\rangle}
\newcommand{\pa}{\|}

\newcommand{\e}{{\rm e}}

\renewcommand{\imath}{\mathrm{i}}

\newtheorem{lem}{Lemma}[section]
\newtheorem{prop}{Proposition}[section]
\newtheorem{thm}{Theorem}[section]

\theorembodyfont{\normalfont\upshape}
 \newtheorem{rem}{Remark}[section]
 \newtheorem{defi}{Definition}[section]
 \newtheorem{ass}{Assumptions}[section]
\makeatletter

\def\intr{\@ifnextchar[\@@intr\@intr}
\def\@@intr[#1]#2{\int_{\rz^{#1}}\!\d^{#1} #2\;}
\def\@intr#1{\int_{\rz}\!\d #1\;}

\DeclareRobustCommand{\qed}{%
  \ifmmode % if math mode, assume display: omit penalty etc.
  \else \leavevmode\unskip\penalty9999 \hbox{}\nobreak%
  \fi
  \hspace{1.2em}\hbox{\qedsymbol}}
\newcommand{\qedsymbol}{\rule[-.4ex]{.9ex}{2ex}}
\newenvironment{proof}[1][\proofname]{\par
  \normalfont
  \topsep6\p@\@plus6\p@ \trivlist
  \item[\hskip10.4mm\sffamily\slshape
    #1\@addpunct{.}\hskip.8em]\ignorespaces
}{%
  \qed\endtrivlist
}
\newcommand{\proofname}{Proof}

\makeatother

\begin{document}

\title{\vspace*{-2cm} Dynamical properties of random
  Schr\"odinger operators} 
\author{Jean-Marie Barbaroux\footnote{e-mail:
  Jean-Marie.Barbaroux@math.univ-nantes.fr} \\ 
  UMR 6629 du CNRS \\ 
  Universit\'e de Nantes\\
  F-44322 Nantes Cedex 3, France. 
  \and  
  Werner Fischer\footnote{ 
  New address: Infineon Technologies, 
  Balanstra\ss{}e 73, D-81541 M\"unchen, Germany}  \\ 
  Institut f\"ur Theoretische Physik \\ Universit\"at
  Erlangen-N\"urnberg\\ Staudtstra\ss{}e 7\\ 
  D-91058 Erlangen, Germany. 
  \and 
  Peter M\"uller\footnote{e-mail:
  Peter.Mueller@Physik.Uni-Goettingen.DE} \\ 
  Institut f\"ur Theoretische Physik \\ 
  Georg-August-Universit\"at \\ 
  D-37073 G\"ottingen, Germany. 
  }

\date{\vspace*{.2cm} \today}

\maketitle

\vspace{-.6cm}
\begin{abstract} 
  We study dynamical properties of random Schr\"odinger operators
  $H^{(\omega)}$ defined on the Hilbert space $\ell^2(\bbZ^d)$ or
  $L^2(\bbR^d)$. Building on results from existing multi-scale
  analyses, we give sufficient conditions on $H^{(\omega)}$ to obtain
  the vanishing of the diffusion exponent 
  $$
  \sigma_{\rm diff}^+ := \limsup_{T\rightarrow\infty } \frac{\log
    \bbE \left(\la\la\vert X
    \vert^2\ra\ra_{T,f_I(H^{(\omega)})\psi}\right) }{\log T}=0.
  $$
  Here $\bbE$ is the expectation over randomness, $f_{I}$ is any
  smooth characteristic function of a bounded energy-interval $I$ and
  $\psi$ is a state vector in the domain of $H^{(\omega)}$ with
  compact spatial support. The quantity $\la\la |X|^2
  \ra\ra_{T,\varphi}$ denotes the Cesaro mean up to time $T$ of the
  second moment of position $\la |X|^2\ra_{t,\varphi}$ at times $0\le
  t\le T$ of an initial state vector $\varphi$. 
  If the Hilbert space is $\ell^2(\bbZ^d)$, the method of proof can be
  strengthened to yield dynamical localization.
  Under weaker assumptions, we also prove a theorem on the absence of
  diffusion. The results are applied to a randomly perturbed periodic
  Schr\"odinger operator on $L^2(\bbR^d)$, to a simple Anderson-type
  model on the lattice and to a model with a correlated random
  potential in continuous space.
\end{abstract}

\vspace{.5cm}
\noindent Key-Words: Random Schr\"odinger operators, diffusion
exponents, absence of diffusion, dynamical localization, correlated
random potentials.

%%% Local Variables: 
%%% mode: latex
%%% TeX-master: "newdyn"
%%% End: 

\section{Introduction} 
%%%%%%%%%%%%%%%%%  INTRODUCTION   %%%%%%%%%%%%%%%%%%%%%%
The study of random Schr\"odinger operators has a long history going
back to the fundamental work of Anderson \cite{anderson} in 1958. 
Random Schr\"odinger operators occur in probabilistic single-particle
models which are commonly accepted \cite{ShEf84,LiGr88} to provide a
minimal description 
for electronic properties of disordered materials such as doped
semiconductors or metals with impurities.
Anderson argued that the presence of disorder induces the so-called
phenomenon of localization: an electron initially located in a bounded
region  will essentially remain there for all times. This, in turn,
should imply a vanishing conductivity -- a fact which is
experimentally verified.

The first rigorous works on random Schr\"odinger operators are due to 
Pastur \cite{pa} who concentrated on their spectral properties. Later
on, these results were extended to establish the 
almost-sure decomposition of the spectrum into pure-point and continuous
parts, to show the existence and regularity properties of the
integrated density of states, as well as exponential localization  
\cite{Kr2,CaLa90,FP}. Here, exponential localization means that
the spectrum is almost surely pure point in some set of energies, with
exponentially decaying eigenfunctions. For energies at the bottom of
the spectrum or near band edges, this property is known for multi-dimensional
Anderson models on the lattice and for certain random Schr\"odinger
operators in multi-dimensional continuous space,
see e.g.\ \cite{KS,FS,holdenmartinelli,vondreifusklein1,AM,CH1,Kl1,BCH,KSS1}. 

Rigorous studies of transport coefficients  
can be found in \cite{KS,FS,holdenmartinelli,BvES,BS}. 
A relevant observable describing transport properties 
of a probabilistic single-particle model is the energy-resolved mean
diffusion constant. For a random Schr\"odinger operator 
$H^{(\omega)}$ on the lattice and an initial state localized at 
$y\in\bbZ^{d}$, it is defined as \cite{holdenmartinelli, BS}
$$
  \lim_{\eps\rightarrow 0}\bbE\left(\eps^2
  \sum_{x\in\ell^2(\bbZ^d)}(x-y)^2 \int_0^\infty {\rm e}^{\eps t}
  \vert \la \delta_x, {\rm e}^{-\imath t H^{(\omega)}} \chi_I({H}^{(\omega)})
  \delta_y\ra\vert^2 dt \right) \ .
$$
Here $\bbE$ is the expectation on the underlying probability space 
$(\Omega, {\cal F},\bbP)$ and $\chi_I$ is the characteristic function
of the set $I\subset\bbR$ of allowed energies.
Other quantities of great interest (see e.g. \cite{AG,BS}), and
closely connected to the previous one, are the diffusion exponents
\begin{eqnarray*}
\sigma_{\rm diff}^- &:=& \liminf_{T\rightarrow\infty } \frac{\log
    \bbE \left(\la\la\vert X \vert^2\ra\ra_{T,
    \chi_I({H}^{(\omega)})\psi}\right) }{\log T}\\
\sigma_{\rm diff}^+ &:=& \limsup_{T\rightarrow\infty } \frac{\log
    \bbE \left(\la\la\vert X \vert^2\ra\ra_{T,
    \chi_I({H}^{(\omega)})\psi}\right) }{\log T},
\end{eqnarray*}
where $\la\la\vert X\vert^2\ra\ra_{T, \chi_I({H}^{(\omega)})\psi}
:= T^{-1}\int_0^T \la\vert X\vert^2\ra_{t,
  \chi_I({H}^{(\omega)})\psi}dt$ is the Cesaro mean of the second moment
of position $\la\vert X\vert^2\ra_{t,
  \chi_I({H}^{(\omega)})\psi} := \pa\,\vert X\vert
\e^{-\imath t H}\chi_I(H^{(\omega)})\psi \pa^2$ 
at times $0\le t\le T$ of an initial state $\psi$ with energy in
$I\subset\bbR$. Until the 
beginning of the 90's, it was generally believed that the occurrence
of pure-point spectrum was a sufficient criterion for the vanishing of
the diffusion exponents
\begin{equation}
 \sigma_{\rm diff}^-=\sigma_{\rm diff}^+=0 \,.\label{introdynloc}
\end{equation}
However, we now know \cite{dRJLS,BCM} that pure-point spectrum and
even exponential localization are not sufficient conditions for
getting ~\eqref{introdynloc} or the vanishing of the diffusion
constant, which follows already from $\sigma_{\rm diff}^+ <1$.
The study of dynamical properties requires additional investigations.
This was done recently by Aizenmann and Graf \cite{AG} who proved the
dynamical localization property
\begin{eqnarray*}
\bbE\left( \sup_{T>1} \,\la\la\vert X \vert^2\ra\ra_{T,
\chi_I({H}^{(\omega)})\psi}
 \right) < \infty\
\end{eqnarray*}
for a large class of discrete random models, i.e. defined on ${\cal
H}=\ell^2(\bbZ^d)$. More recently, Germinet and De Bi\`evre 
\cite{debievregerminet} showed $\sup_{T>0}\la\vert X
\vert^2\ra_{T, \chi_I({H}^{(\omega)})\psi} < \infty$, $\bbP$-almost
surely, for both discrete and continuous models exhibiting exponential
localization in $I$. This is quite a strong result, since it gives a
bound for a quantity which has not been averaged over time. However,
this bound provides no information about the expectation $\bbE$.  Our
main goal here is to show that the vanishing of diffusion exponents
\eqref{introdynloc} also holds for a large class of random
Schr\"odinger operators on $\ell^2(\bbZ^d)$ or $L^2(\bbR^d)$ if we
replace the characteristic function $\chi_I$ by any smooth function
$f_{I}$ with compact support in $I$ (Theorem~\ref{maintheorem}).  
For Schr\"odinger operators on $\ell^2(\bbZ^d)$ the method of proof
can be strengthened to yield dynamical localization
(Theorem~\ref{maindiscrete}). By
relaxing some of the assumptions, a result on the absence of diffusion
is also established (Theorem~\ref{maintheorem2}).

Similar to the strategy of Fr\"ohlich and Spencer \cite{FS}, who
related the conductivity at some fixed energy $E$ to the behavior of
the Green's function at this energy, it is the strategy of the present
paper to derive \eqref{introdynloc} from appropriate decay estimates
of the Green's function. However, in order to calculate $\bbE\left(
\la\la\vert X \vert^2 \ra\ra_{T, f_{I} (H^{(\omega)} ) \psi} \right)$ we
need a more refined analysis that takes into account the dependence of
the Green's function on the energy $E$ and on the realization
$\omega\in\Omega$ of the random potential. This can be done with the
help of von Dreifus and Klein's estimates \cite{vondreifusklein1} on
the decay of the Green's function which are uniform in energy (see
\eqref{msaAss} below).  For proving absence of diffusion, it suffices
to have fixed-energy decay estimates
(see \eqref{msaAss2} below). Although the second result on
the vanishing of the diffusion constant is weaker than the one of vanishing
diffusion exponents, it is worth to be mentioned, since the required
fixed-energy 
estimates are proven for a very large class of random operators.  On
the contrary, decay estimates which are uniform in energy are only
known for fewer cases (see e.g. \cite{vondreifusklein1, Kl1,
debievregerminet, KSS1}).

The paper is organized as follows: In Section~$2$ we present the
assumptions needed and we state our main results. In Section~$3$ we
consider some examples to illustrate the applicability of our theorems.
Periodic continuous Schr\"odinger operators perturbed by an alloy-type
random potential serve as applications for both 
Theorem~\ref{maintheorem2} on the absence of diffusion and for
Theorem~\ref{maintheorem} on vanishing diffusion exponents
at energies near band-edges.
Example~2 is concerned with an application of
Theorem~\ref{maindiscrete}  and
establishes dynamical localization for a 
multi-dimensional Anderson model on the lattice.
Finally, Example~3 deals with a Schr\"odinger operator with a
correlated alloy-type random potential. We show that the diffusion
exponents of this model get smaller and smaller 
for energies approaching the bottom of the spectrum.
For this purpose we need a ``variable-energy'' multi-scale analysis that gives
algebraic decay estimates which are uniform in energy.  Details of
these calculations are deferred to the Appendix. Section 4 is devoted
to the proofs of the theorems.

%%% Local Variables: 
%%% mode: latex
%%% TeX-master: "newdyn"
%%% End: 

\section{Main results}
We first describe precisely the quantities of interest. For the
physical relevance of these quantities, one can refer for example to
\cite{thouless, BvES, AG, BS}. 

%%%%%%%%%%%%%%%%%%%%%%%%%%%%%%%%%%%%%%%%%%%%%%%%%%%%%%%%%%%%%
%%%%% DEFINITION 2 Moment of order 2 / Diffusion exponent %%%
%%%%%      localization length / diffusion exponent       %%%
%%%%%%%%%%%%%%%%%%%%%%%%%%%%%%%%%%%%%%%%%%%%%%%%%%%%%%%%%%%%%
Consider a probability space $(\Omega, {\cal F}, \bbP)$ and a random
operator $H$, that is a random variable $\Omega\ni\omega\mapsto
H^{(\omega)}$ which takes on values in the space of linear operators
on a Hilbert space $\calH$.  Here $\calH$ will either be the space
$\L2(\bbR^{d})$ of square-integrable functions or the space $
\ell^2(\bbZ^{d})$ of square-summable sequences.
By $X=(X_1, \ldots , X_d)$ we denote the self-adjoint multiplication
operator on $\calH$, and $|X|:=\left( \sum_{j=1}^{d}
X_j^{2}\right)^{1/2}$ is its modulus. For the random operator $H$ we
consider the following
%
%%%%%%%%%  ASSUMPTIONS   %%%%%%%%%%%%%
\begin{ass} \label{ass}
~\\[-4mm]
\begin{itemize}
\item[i)] $H$ is measurable
  %ergodic (in the sense of \cite{Kr2} and \cite{FP})
  and $\bbP$-almost surely self-adjoint with constant domain
  $\calD(H) $.
\item[ii)]
  There is $\calC\subseteq \calH $ such that $\calC$ is
  $\bbP$-almost surely a core for $H$ and 
  $ (\imath + \varepsilon X_j )^{-1} \calC \subseteq \calC $ for all
  $\varepsilon \in \, ]0,1[ \, $ and $ j=1, \ldots , d$.
\item[iii)]
  The commutators $ [H,X_j]:=HX_j-X_j H $, $j=1, \ldots , d$, are 
  well-defined and relatively operator bounded with respect to $H$,
  i.e. for $\bbP$-a.e.\ $\omega\in\Omega$ there exists $b(\omega)<\infty$
  such that $\pa [H^{(\omega)}, X_j](H^{(\omega)}+\imath)^{-1} \pa \leq
  b(\omega)$. We further assume that 
  $\bbE\bigl(b^2(\omega)\bigr)<\infty$. 
\item[iv)] $\calD(H)\cap\calD(\vert X \vert)$ is dense in $\calH$.
\item[v)] For $L>1$, $ z\in \bbC\setminus\bbR$  and $q,q'\in
  \bbZ^d$ with $ \pa q-q' \pa_\infty > 2L$ there is an operator $
  D^{(\omega)}_{L,q'}(z)$ defined on $\calD(H^{(\omega)})$ 
  %$ \one_{q'} \calH \subseteq\calD(D^{(\omega)}_{L,q'}(z)) $ 
  such that
  \begin{equation}
    \label{resEqn}
    \one_{q'} (H^{(\omega)}-z)^{-1} \one_q =
    \one_{q'} D^{(\omega)}_{L,q'}(z) (H^{(\omega)}-z)^{-1} \one_q
  \end{equation}
  holds for $\bbP$-almost every $\omega$. Here $ \one_q $ acts as the
  multiplication operator corresponding to the function
  \begin{equation}\label{charac}
    \one_q(x):=\left\{
      \begin{array}{rl}
        1 & \text{ if } \pa x-q\pa_\infty < 1 \, ,\\
        0 & \text{ else }\, . 
      \end{array}
    \right.
  \end{equation}
\end{itemize}
\end{ass}
%%%%%%%%%%%%%%%%%%%%%%%%%%%%%%%%%%%%%%%%%%%%%%%%
%
%%%%%%%%%%%%%%%%%  REMARK %%%%%%%%%%%%%%%%%%%%%%
\begin{rem}\label{rem1}
  For conditions on the random operator $H$ to fulfill i), see
  \cite{Kr2,CaLa90,FP}. Among others, the measurability of $H$ assures
  that functions of the type
  $(E,\omega) \mapsto \pa (H^{(\omega)}-E-\imath\eps)^{-1}\varphi\pa$,
  which are continuous in $E$, are
  jointly measurable in $E$ and $\omega$. Assumption ii) is true
  for a large class of operators with
  $\calC=C_0^\infty(\bbR^d)$, see e.g.  \cite{HS}. Assumption iii)
  naturally occurs in the proof of Lemma~\ref{heisenberg} and helps to
  guarantee the finiteness, for all strictly positive $T$,
  of the quantity $\la\la \vert X \vert^2\ra\ra_{T,\psi}$ given below in
  Definition~\ref{defmoment}. Assumption v) assures the existence of a
  geometric resolvent equation, which is written down in an abstract form
  in Eq.\ \eqref{resEqn} in order to suit both the
  discrete and the continuous case. 
  Concrete examples for the operators $D^{(\omega)}_{L,q'}(z)$, which
  in the sequel
  will be supposed to exhibit some nice decay properties, can be found
  in Section~\ref{S3} (see also \cite{CH1} and \cite{vondreifusklein1}).
\end{rem}
%%%%%%%%%%%%%%%%%%%%%%%%%%%%%%%%%%%%%%%%%%%%%%%%
%
%%%%%%%%%%%%  DEFINITION 1 - ( )-regular   %%%%%
\begin{defi}
Let $\rho : \bbR_+ \to \bbR_+ $ be a non-negative function.  The
operator $H^{(\omega)} $ is said to be $(\rho,E,L,q)$-regular, if
$$
  \sup_{\varepsilon \not= 0} \pa \one_{q} D^{(\omega)}_{L,q}(E+\imath
  \varepsilon ) \pa \le \rho(L)^{1/2} \, .
$$
\end{defi}
%%%%%%%%%%%%%%%%%%%%%%%%%%%%%%%%%%%%%%%%%%%%%%%
%\begin{rem}\label{rem1bis}
%  Equation \eqref{resEqn} mainly refers to the geometric
%  resolvent equations, in the discrete case as well as in the
%  continuous case.  Details can be found in Section~\ref{S3} (see also
%  \cite{CH1} and \cite{vondreifusklein1}).
%\end{rem}
%
%%%%% MULTI-SCALE ASSUMPTION for vanishing diff. exponent  %%%%%%%%%%%%
\textbf{Multi-Scale Assumption 1}
\begin{itemize}
\item[(M1)] For $H$ satisfying Assumptions \ref{ass}.i) and v), we
  suppose that 
  there exist $L_0>1$, $\alpha>1$, $\nu>0$, $p>\alpha (d+2)$, a
  bounded interval $I\subset \bbR$ and a non-negative function
  $\rho : \bbR_{+} \to \bbR_+ $ such that for
  $L_k:=L_0^{\,\alpha^k}$ one has
$$
\rho(L_k)\leq \exp\{-2L_k^{\,\nu}\}\ ,
$$
and
\begin{multline}\label{msaAss} 
\bbP\bigl\{ \omega\ \vert\ H^{(\omega)}\text{ is }
  (\rho,E,L_k,q'')-\text{regular for all } E \in I \\ \text{ and
  for either } q''=q \text{ or } q''=q' \bigr\}\ge 1-L_k^{-p} ,
\end{multline}
for all $k\in\bbN$ and all $q,q'\in \bbZ^d$ with 
$\pa q-q'\pa_\infty >2 L_k$.
\end{itemize}

Losely speaking, (M1) assumes a decay with good probability of the
localized resolvents (see the examples in Section \ref{S3}). Such a
decay can be shown by performing a ``variable-energy'' multi-scale
analysis in the spirit of von Dreifus and Klein (see
e.g. \cite{vondreifusklein1,Kl1,debievregerminet}).  For the weaker
Theorem~\ref{maintheorem2} to hold, we only need a weaker form of
(M1), in that the set of events giving rise to decay need not be
uniform in energy.

%%%%%%%%%  MULTI-SCALE ASSUMPTION for absence of diffusion  %%%%%%
\medskip
\noindent 
\textbf{Multi-Scale Assumption 2}\nopagebreak
\begin{itemize}
\item[(M2)] For $H$ satisfying Assumptions \ref{ass}.i) and v), we
  assume that 
  there exist $L_0>1$, $\alpha>1$, $\beta>0$, $m>(3+d+\beta)$,
  $p>3+d+\beta$ such that for $\rho(x):=x^{-m}$,
  $L_k:=L_0{}^{\alpha^k}$ and for all $E\in I$ and all $q\in\bbZ^d$, $\pa q
  \pa_\infty> 2 L_k$, we have
\begin{eqnarray}\label{msaAss2}
  \bbP\left\{ \omega\ \vert\ H^{(\omega)}\text{ is } 
  (\rho,E,L_k,q)-\text{regular} \right\}\ge 1-L_k^{-p}\ .
\end{eqnarray}
\end{itemize}

\begin{defi}\label{defmoment}
Take $\psi\in{\cal H}$ such that for all $t>0$, $\e^{-\imath t
H}\psi\in{\cal D}(\vert X\vert )$. The Cesaro mean up to time $T>0$ of
the second moment of $\psi$ is defined as
$$
    \la\la\vert X \vert^2 \ra\ra_{T,\psi} := \frac{1}{T}\int_0^T
    \pa\,\vert X\vert \e^{-\imath t H}\psi \pa^2 dt \ ,
$$
where $\imath$ is the imaginary unit. By analogy with \cite{AG} and \cite{BS}
we say that $\psi$ is dynamically localized if
$$
    \bbE\left( \sup_{T>1}\, \la\la\vert X \vert^2
    \ra\ra_{T,\psi}\right)<\infty\ ,
$$
where $\bbE$ is the expectation associated with $\bbP$. The diffusion
exponents are defined as in \cite{BS}
\begin{eqnarray*}
    \sigma^+_{{\rm diff}}(\psi) & := & \limsup_{T\rightarrow\infty}
    \frac{\log \bbE\left( \la\la\vert X \vert^2 \ra\ra_{T,\psi}\right)
    }{\log T}\ , \label{expodiff}\\ 
    \sigma^-_{{\rm diff}}(\psi) & := &
    \liminf_{T\rightarrow\infty} \frac{\log \bbE\left( \la\la\vert X
    \vert^2 \ra\ra_{T,\psi}\right) }{\log T}\ .
\end{eqnarray*}
When the limit exists, the diffusion constant is given by
$$
    D(\psi):=\lim_{\eps\rightarrow 0} \bbE\left(\eps^2 \int_{\bbR}
    \pa\,\vert X \vert (H^{(\omega)} - E -\imath\eps)^{-1} \psi \pa^2\
    dE \right)\ .
$$
\end{defi}
%%%%%%%%%%%%%%%%%%%%%%%%%%%%%%%%%%%%%%%%
%
\begin{rem}
  If the upper diffusion exponent obeys $\sigma^+_{{\rm diff}}(\psi) <
  1$, then the diffusion constant vanishes, $D(\psi)=0$.
\end{rem}
%
%%%%%%%%%%%%%%%%%%%%%%%%%%%%%%%%%%%%%%%%

The main results of this paper are summarized in the following 
theorems.
%%%%%%%%%%%%%%%%%%%%%%%%%%%%%%%%%%%%%%%%%
%
%%%%%%%    THEOREM    1  Vanishing Diffusion Exponents   %%%%%%%%%%%%%%
\begin{thm}\label{maintheorem} (Vanishing diffusion exponents).\\
Consider a random operator $H$ satisfying Assumptions \ref{ass} and
the Multi-Scale Assumption (M1).  Let $I'$ be any compact subset of
$I$ such that ${\rm dist}(\partial I, \partial I') >0$. Then for all
compactly supported $\varphi \in \calD(H)$ and all $f_{I'} \in
C_0^{\infty}(I') $ one has
$$ 
   \sigma^{\pm}_{{\rm diff}}\bigl(f_{I'}(H)\varphi\bigr)=0\,.
$$
This implies in particular that the diffusion constant is zero,
$D\bigl(f_{I'}(H)\varphi\bigr)=0$.
\end{thm}

\begin{rem}\label{rem2.3}
If the decay of the function $\rho$ is only algebraic,
one can still obtain an estimate for the diffusion exponents. Namely
if there exists $n>\alpha(d+2)$ and $c(n)<\infty$ such that
$$
\forall k\in
\bbN, \ \ \rho(L_k) < c(n) L_k^{-2n}\ ,
$$
then we get according to Remark~\ref{remdynloc}.ii)
$$
 \sigma^-_{{\rm diff}}\bigl(f_{I'}(H)\varphi\bigr)\leq \sigma^+_{{\rm
 diff}}\bigl(f_{I'}(H)\varphi\bigr)\leq\frac{\alpha(d+2)}{n} .
$$
\end{rem}

For discrete random Schr\"odinger operators a stronger result is
stated in 
%%%%%%%%%%%%%%%%%%%%%%%%%%%%%%%%%%%%%%%%%%%%%%%%%%%%%%%%%%%%%%%%
%
%%%%%%%    THEOREM    2  Dynamical localization   %%%%%%%%%%%%%%
\begin{thm} \label{maindiscrete} (Dynamical localization).\\ 
Suppose that the assumptions of Theorem~\ref{maintheorem} are
fulfilled and that the Hilbert space is
${\cal H}=\ell^2(\bbZ^d)$. Then one has dynamical localization
\begin{equation}\label{13bis}
  \bbE\left( \sup_{T>1}\, \la\la\vert X \vert^2
  \ra\ra_{T,f_{I'}(H)\psi}\right)<\infty\,.
\end{equation}
\end{thm}

In case one can only establish the weaker Multi-Scale Assumption (M2),
a weaker dynamical property can still be shown.
%%%%%%%%%%%%%%%%%%%%%%%%%%%%%%%%%%%%%%%%%%%%%%%%%%%%%%%%%%%%%%%%%%%%%
%
%%%%%%%%%%%%%%%%%  THEOREM 3 Absence of diffusion  %%%%%%%%%%%%%
\begin{thm}\label{maintheorem2} (Absence of diffusion).\\
Let $H$ be a random operator satisfying Assumptions \ref{ass} and the
Multi-Scale-Assumption (M2), and let $I^\prime$ be as above.  Then for
all compactly supported $\varphi \in \calD(H)$ and all $ f_{I'} \in
C_0^{\infty}(I') $, we have
$$
    \sigma^+_{{\rm diff}}\bigl(f_{I'}(H)\varphi\bigr) < 1\ .
$$
This implies in particular that the diffusion constant vanishes,
$D\bigl(f_{I'}(H)\varphi\bigr)=0$.  \end{thm}

%%% Local Variables: 
%%% mode: latex
%%% TeX-master: "newdyn"
%%% End: 

\section{Applications}\label{S3}
We present here some examples of random Schr\"odinger operators for
which the assumptions of one of the above theorems 
are fulfilled. The first example concerns both absence
of diffusion and vanishing diffusion exponents near
band edges. The second example concerns dynamical
localization and the third one 
smaller and smaller diffusion
exponents for energies approaching the bottom of the spectrum.  Our
aim is to use, as much as possible, the spectral results already known
for these models in order to prove either (M1) or (M2), thus showing
that our theorems can easily be applied to a very large class of
random Schr\"odinger operators. 

We will adopt the following notations:
For $L>0$ and $x\in\bbR^d$, $\Lambda_L(x)$ is the cube $\{y\in\bbR^d \ 
\vert\ \pa x-y \pa_\infty \leq L \}$ and $\chi_{\Lambda_L(x)}$ denotes
a smooth characteristic function of $\Lambda_L(x)$, i.e.
$\chi_{\Lambda_L(x)}\in C^2(\bbR^d)$ and for some fixed $\delta>0$,
\begin{eqnarray*}
  \chi_{\Lambda_L(x)}= 
  \begin{cases}
    1  & \ {\rm if} \ \ \pa x-y\pa_\infty \leq L-\delta\ ,\\
    0  & \ {\rm if} \ \ \pa x-y\pa_\infty \geq L\ ,\\
    \in[0,1] &\ {\rm elsewhere}\ .
  \end{cases}
\end{eqnarray*}
%%%%%%%%%%%%%%%%%%%%%%%%%%%%%%%%%%%%%%%%%%%%%%%%
%%%%%%%%%%    Example 1   %%%%%%%%%%%%%%%%%%%%%%
%%%%%%%%%%%%%%%%%%%%%%%%%%%%%%%%%%%%%%%%%%%%%%%%
\textbf{Example 1: Random perturbations of periodic continuous
  Schr\"o\-din\-ger operators}

We consider a specific case of the random Schr\"odinger operators
studied in \cite{BCH},
$$
H^{(\omega)} = -\Delta + V_{\rm per} + V^{(\omega)}\ \ {\rm on}\
L^2(\bbR^d)\ ,
$$
where $V_{\rm per}$ is a bounded periodic potential such that
$-\Delta + V_{\rm per}$ has a gap $(B_-,B_+)$ in its spectrum and
$V^{(\omega)}=g \sum_{i\in\bbZ^d} \lambda_i(\omega) u(x-i)$ with

\begin{itemize}
\item[(H1)] The potential $u$ is positive, with compact support, such
  that $u(x)\geq 1$ on $[-1/2, 1/2]^{d}$, $\pa u \pa_\infty < \infty$.
\item[(H2)] $(\lambda_i)_{i\in\bbZ^d}$ is a stationary process of
  independent and identically distributed random variables. We assume
  that the probability distribution of $\lambda_i$ has a Lebesgue
  density $h\in C^0(\bbR)$, with compact support $[-M, M]$, and for some
  $\nu>0$, $\bbP\{\mid\lambda\pm M\mid <\eps \}\leq\eps^{d+2+\nu}$.
\item[(H3)]
  \begin{sloppypar}
    The coupling constant $g$ satisfies $g < (B_+ - B_-)/(2 M u_{\rm
      max})$, where ${u_{\rm max}=\pa\sum u(x-i)\pa_\infty}$.
  \end{sloppypar}
\end{itemize}
It is proven in \cite{BCH} that there exist non-empty compact subsets
of $\bbR$, $I_+\neq\emptyset$ and $I_-\neq\emptyset$, at the edge of
the almost-sure spectrum of $H^{(\omega)}$, such that for
$\alpha=(d+4)/(d+1)$, $m=p=4+d$, and for some $L_0<\infty$, if
$L_k=L_0^{\alpha^k}$, $E\in I_+\cup I_-$, $q'\in\bbZ^d$,
$$
  \bbP\{\omega \ \vert\ \sup_{\eps\neq 0} \pa [-\Delta,
  \chi_{\Lambda_{L_k}(q')}] (H_{\Lambda_{L_k}(q')}^{(\omega)} -E
  -\imath\eps)^{-1} \unc_{q'}\pa\leq L_k^{-\frac{m}{2}}  \}\geq
  1-L_k^{-p}\ , 
$$
where $H_{\Lambda_{L_k}(q')}^{(\omega)}=-\Delta + V_{\rm per} +
\sum_{i\in\bbZ^d\cap \Lambda_{L_k}(q')}\lambda_i(\omega) u(x-i)$ and
$\unc_{q'}$ is defined in \eqref{charac}. According to the geometric
resolvent equation, for $L>0$ and $q'\in\bbZ^d$,
\begin{eqnarray*}
  \lefteqn{\chi_{\Lambda_{L}(q')} (H^{(\omega)} - E -\imath\eps)^{-1} 
    =   (H_{\Lambda_L(q')}^{(\omega)} - E
    -\imath\eps)^{-1}\chi_{\Lambda_{L}(q')}\ \ \ \ \ \ \ \ \  } \\ 
  &  & \ \ \ \ \ \ \ \ \ \ + (H_{\Lambda_L(q')}^{(\omega)} - E
  -\imath\eps)^{-1} [-\Delta, \chi_{\Lambda_L(q')}](H^{(\omega)} - E
  -\imath\eps)^{-1} \ .
\end{eqnarray*}
This implies \eqref{resEqn} and (M2) with $D_{L,q'}^{(\omega)}(z) =
(H_{\Lambda_L(q')}-z)^{-1} [-\Delta, \chi_{\Lambda_L(q')} ]$.
Assumptions~\ref{ass}.i) -- v) are easily verified with ${\cal
  C}=C_0^\infty(\bbR^d)$, and $b(\omega)=3(1+\|V_{\rm per}\|_\infty +
(B_+ - B_-)/2)$; thus the conclusion of Theorem~\ref{maintheorem2}
holds for this model with $I=I_+ \cup I_-$.

%%%%%%%%%%%%%%%%%%%%%%%%%%%%%%%%%%%%%%%%%%%%%%%%
%%%%%%%%%%    Example 1  (again) %%%%%%%%%%%%%%%
%%%%%%%%%%%%%%%%%%%%%%%%%%%%%%%%%%%%%%%%%%%%%%%%
\medskip\noindent
\textbf{Example 1 (revisited): Long-range single-site potentials}

The model considered above -- with the coupling constant $g$
set equal to one -- was also studied in
\cite{KSS1} under less restrictive assumptions on the random
potential. Moreover, a ``variable-energy'' multi-scale analysis with
exponential decay estimates was
established there for this model. It allows us to apply
Theorem~\ref{maintheorem} for energies near band edges,
thus yielding vanishing diffusion exponents in this regime.
We shall only be concerned with a particular case of \cite{KSS1} where
Hypotheses (H1) -- (H3) are replaced by
\begin{itemize}
\item[(H1')] The single-site potential $u\ge 0$ obeys $u\ge c>0$ on some
  non-empty open set, is bounded and decays algebraically 
  $u(x) \le C (1+|x|)^{-m}$ with some constants $C<\infty$ and 
  $m>8(d+1)$.
\item[(H2')] $(\lambda_i)_{i\in\bbZ^d}$ is a stationary process of
  independent and identically distributed random variables. We assume
  that the probability distribution of $\lambda_i$ has a bounded Lebesgue
  density with compact support $[-M,M]$ such that
  $\bbP\{\mid\lambda\pm M\mid <\eps \}\leq\eps^{\tau}$ for all small 
  $\eps >0$ and some $\tau > d+1$.
\end{itemize}
As above, Assumptions~\ref{ass} i) -- v) are satisfied with ${\cal
C} = C_0^\infty(\bbR^d)$, with an almost-surely uniformly bounded
$b(\omega)$ (since $V^{(\omega)}=\sum_{i\in\bbZ^d} \lambda_i(\omega)
u(x-i)$ is almost-surely uniformly bounded) and with
$D_{L,q'}^{(\omega)}(z)$ as defined in Example~1. The Multi-Scale
Assumption (M1) is provided by Thm.~4.3 in \cite{KSS1} for energies
near band edges. To see this,
we note that our decay exponent $p$ corresponds to the quantity $2\xi$
in \cite{KSS1}. Thm.~4.3 in \cite{KSS1} holds for
any positive value of $\xi$ subject to the conditions $\xi < 2\tau - d$
and $\xi < (m/4) - d$ (Prop.~3.5(b) and Thm.~4.1 in \cite{KSS1}). Hence, 
(H1') and (H2') guarantee that values $\xi > d +2$ are allowed. 
But since $\alpha < 2$ in \cite{KSS1}, it follows that Thm.~4.3 in
\cite{KSS1} holds with $2\xi > \alpha (d+2)$, as required in (M1).
Thus, we can apply Theorem~\ref{maintheorem} in order to get a
vanishing diffusion exponent for energies near band edges of the
almost-sure spectrum of $H^{(\omega)}$.

Random Schr\"odinger operators with 
single-site potentials $u$, which may also take on negative values,
have  been studied
in \cite{Kl1}. There a ``variable-energy'' multi-scale analysis with
exponential decay is proven for energies near the bottom of the
spectrum. The result enables one to apply Theorem~2.1, thereby
establishing vanishing diffusion exponents under these circumstances,
too.

%%%%%%%%%%%%%%%%%%%%%%%%%%%%%%%%%%%%%%%%%%%%%%%%
%%%%%%%%%%    Example 2   %%%%%%%%%%%%%%%%%%%%%%
%%%%%%%%%%%%%%%%%%%%%%%%%%%%%%%%%%%%%%%%%%%%%%%%

\medskip\noindent
\textbf{Example 2: Discrete Anderson model}

We consider the random family \cite{anderson, FS, vondreifusklein1,
  CaLa90, FP} 
$$
H^{(\omega)}=-\Delta_d + V^{(\omega)}\ \ {\rm in\ }\ell^2(\bbZ^d)\ ,
$$
where $(-\Delta_d \psi)(n):= \sum_{i\in\bbZ^d,\, |i-n|=1}\psi(i)$,
is the discrete Laplacian and $V^{(\omega)}(n)$, $n\in\bbZ^d$, are
independent and identically distibuted random variables with
absolutely continuous density $g(\lambda):= d\mu /d\lambda$ satisfying
$\int \lambda^2 g(\lambda) d\lambda<\infty$. Although dynamical
localization has already been proved in \cite{AG} for this model, we
reconsider this issue to demonstrate the applicability of
Theorem~\ref{maindiscrete}.  

For the discrete Anderson model, Assumptions~\ref{ass}.i), ii) and iv)
are true with $\calC = \calD (H^{(\omega)}) = \calH$.  Since
$[H^{(\omega)},X_j]$ is bounded, Assumption~\ref{ass}.iii) is
fulfilled with $b(\omega)= 2d$ for $\bbP$-a.e. $\omega$. Now, the
usual geometric resolvent
equation in $\ell^2(\bbZ^d)$ gives, for $\Im z\neq 0$, and for all
$q,q'\in\bbZ^d$ such that $\pa q-q' \pa_\infty>2L$,
\begin{eqnarray}
  \lefteqn{\la \unc_{q'},\, (H^{(\omega)}-z)^{-1}\unc_q\ra  }\nonumber \\
  & \!\!\!\! = \sum_{u\in\Lambda,
    u'\not\in\Lambda,\pa u-u'\pa_\infty=1}\la \unc_{q'},\,
  (H_{\Lambda_L(q')}^{(\omega)}-z)^{-1} \unc_u\ra\,\la\unc_{u'},\,  
  (H^{(\omega)}-z)^{-1}\unc_q\ra\label{discreteGRE}\ ,
\end{eqnarray}
where $H_{\Lambda_L(q')}^{(\omega)}$ is the operator $H^{(\omega)}$
restricted to the box $\Lambda_L(q')$ with Dirichlet boundary
conditions. Thus \eqref{discreteGRE} gives us the property v) for $H$,
if we define $D^{(\omega)}_{L,q'}(z)$ by its action on $\varphi\in{\cal
  H}$ according to 
\begin{eqnarray}
  \left(D_{L,q'}^{(\omega)}(z)\varphi\right)(x)=
  \sum_{u\in\Lambda,
    u'\not\in\Lambda,\pa u-u'\pa_\infty=1}\la\unc_{u'},\,\varphi\ra
  (H_{\Lambda_L(q')}^{(\omega)}-z)^{-1}\unc_{u}(x) \ .\nonumber 
\end{eqnarray}

Moreover, thanks to \cite[Theorem 2.2 and Proposition
A.11]{vondreifusklein1}, there exists $0<E_0<\infty$ such that for
$I= (-\infty, -E_0]\cup [E_0,+\infty)$ and for all $q,q'$, $\pa
q-q'\pa_\infty >2L$, we have
\begin{eqnarray}
  \bbP \Big\{\omega\vert\ \forall E\in I,\, 
  H^{(\omega)} {\rm is\ } (\rho,
  E, L_k,q''){\rm-regular\ for\ either\  }
  q''=q \  {\rm or\  }q''=q'\Big\}  & & \nonumber \\
  \!\!\!\!  \geq 1-L_k^{-p}\ ,& & \nonumber
\end{eqnarray}
with $p>4d+2$, $\rho(x)=\e^{-mx/2}$ for some fixed $m>0$,
$L_k=L_0^{(3/2)^k}$ for $L_0$ finite depending only on $m$ and $p$.
This gives exactly the Multi-Scale Assumption (M1). One can thus apply
Theorem~\ref{maindiscrete} with any compact subset $I'$ of $I$.

%%%%%%%%%%%%%%%%%%%%%%%%%%%%%%%%%%%%%%%%%%%%%%%%
%%%%%%%%%%    Example 3   %%%%%%%%%%%%%%%%%%%%%%
%%%%%%%%%%%%%%%%%%%%%%%%%%%%%%%%%%%%%%%%%%%%%%%%
\medskip\noindent
\textbf{Example 3: Anderson model with correlated potentials}

We consider the Hamiltonian on $L^2(\bbR^d)$
\begin{eqnarray}\label{model2}
  H^{(\omega)}=-\Delta + \sum_{i\in\bbZ^d}\lambda_i(\omega) u(x-i)\,, 
\end{eqnarray}
for which the random potentials $\lambda_i(\omega)u(x-i)$ at each site
$i$ are correlated. This model has been studied e.g. by
\cite{vondreifusklein2, CHM1}. Correlated potentials means here that
the coupling constants $\lambda_i$ are not independently distributed.
In this case, it has been proven that the bottom of the spectrum is
dense pure point. Our goal here is to prove, with the help of a
``variable-energy'' multi-scale analysis, dynamical localization at
the bottom of the spectrum. Let us first present the assumptions for
this model:

\begin{itemize}
\item[(A1)]
  The site potential $u(x)$ is non-negative, not identically zero, compactly
  supported and $\pa u \pa_{\infty}<\infty$.
\item[(A2)]
  $\{ \lambda_i\}_{i\in\bbZ^d}$ forms a stationary stochastic process of
  identically distributed random variables. 
\item[(A3)] The conditional probability distribution of $\lambda_0$,
  given $\lambda^\perp_0=\{ \lambda_i,\ i\neq 0\}$, is absolutely
  continuous with respect to Lebesgue measure with density $h_0$ which
  is compactly supported, $\pa h_0 \pa_\infty < \infty$ and
  $\bbE(|\lambda_0|^{\max(2,d)})<\infty$.
\item[(A4)] Let $A$ be any given event on $\Lambda_L(0)$ (i.e.
  depending only on $\lambda_i(\omega)$,
  $i\in\Lambda_L(0)\cap\bbZ^d$). We denote by $A(x)$ the event $A$
  shifted to $\Lambda_L(x)$. For any given $\alpha>1$ and $\beta>1$,
  we assume that there exist $K_0(\alpha)$ even and $C(K_0)<\infty$
  such that for all integer $K\geq K_0$, one can find
  $\theta(K,\alpha)> \beta$ having the property that for $L\gg 1$,
  $\forall x_1,x_2,\cdots,x_K\in\bbZ^d$ with $\pa
  x_i-x_j\pa_\infty\geq (1/2) L^\alpha$, ($i\neq j$), and for all
  events A on $\Lambda_L(0)$,
  $$
  \bbP \Big\{\cap_{i=1}^K A(x_i)   \Big\} \leq C\,\bbP(A)^\theta\ .
  $$ 
\end{itemize} 
Existence of random processes satisfying (A2) -- (A4) are given for
example in \cite{vondreifusklein2}. Under these hypotheses, it is
known that $H$ is an ergodic family of almost-surely essentially
self-adjoint operators on $C_0^\infty(\bbR^d)$. 
This implies i), ii) and iv) of
Assumption~\ref{ass}. Furthermore, for $\bbP$-a.e. $\omega$, the
potential $V^{(\omega)}$ is uniformly bounded in $\omega$, and thus
$[H^{(\omega)}, X_j]$ is relatively $H^{(\omega)}$-bounded, with
relative bound $2(1 + \pa V \pa_\infty^2)^{1/2}$. 
Assumption~\ref{ass}.v) is simply the geometric resolvent equation,
i.e., for $1\leq 
L<\infty$ and $q$, $q'$, such that $\pa q - q'\pa_\infty > 2L$ one
has,
$$
  \unc_{q'}(H^{(\omega)}-z)^{-1}\unc_{q}  = \unc_{q'}
  (H^{(\omega)}_{\Lambda_L(q')} - z)^{-1} [H^{(\omega)},
  \chi_{\Lambda_L(q')} ] (H^{(\omega)} - z )^{-1} \unc_q\ ,
$$
where $\chi_{\Lambda_L(q')}$ is precisely defined in
\eqref{caracteristic}. Let $E_- := \inf({\Sigma})$ denote the
bottom of the almost-sure spectrum of $H^{(\omega)}$. Following
\cite{CH1, vondreifusklein2}, we know that for all $p>0$, $m>0$, $S$,
$N$, and $L_0 <\infty$ there exists $E(L_0)>E_-$ such that with the
notations of the Appendix and $I=[E_-, E(L_0)]$, we have the initial
decay:
\begin{multline}
  \bbP\left\{  \forall E\in I,\ \exists (n'_i)_{i=1,\dots S}\ {\rm s.
      t.\ } 
    \gH_{L_1}(E,y_1,m,(n'_i)) \ 
    \ {\rm or\ }   
    \gH_{L_1}(E,y_2,m,(n'_i)) \  \right\} \\
  \geq  1-L_0^{-p}\ , \nonumber
\end{multline}
where $\gH_{L_1}$ is a decay property for resolvents precisely defined
in \eqref{equ.b} of the Appendix. Thus, according to the estimate
\eqref{prob3} and Lemmas~\ref{deterministiclemma} and
~\ref{probabilisticlemma} we take $\alpha=3/2$ and $K_0>3d+5$ such that
$\theta(K_0, 3/2)>3/d+1/2$; We also fix $S=4$, $N=4+K_0$,
$p=(K_0+1)/2$, $w=2d+K_0/2 +1$ and $m>4w+2(d-1)$.  Then we obtain,
with the notations of the Appendix, for $L_k=L_0^{ (\frac{3}{2})^k }$,
and $q$, $q'$ such that $\pa q - q' \pa_\infty > 2L_k$,
\begin{eqnarray}
  \lefteqn{ \bbP \Big\{ \forall E\in I \ {\rm either\ for\ } \tilde{q}=q\
    {\rm or}\ \tilde{q}=q',\  
    }\nonumber \\
  & & \!\!\!\!\!\sup_{\eps>0} \pa [-\Delta,\, \chi_{N}^{L_k}(\tilde{q})]
  (H_{\Lambda_{L_k(\tilde{q})}} -E 
  -\imath\eps)^{-1} \chi_0^{L_k}(\tilde{q}) \pa \leq L_k^{-m} \Big\}\!\geq\!
  1-L_k^{-p}\ .\label{alldecay}
\end{eqnarray} 
This inequality is not the same as the one required in equation
~\eqref{msaAss} of Assumption (M1) since the decay of resolvents is
only algebraic instead of being exponential. However, $m$ can
be chosen larger and larger, if the interval $I$ of energies gets
closer and closer to the bottom of the spectrum. Therefore, according
to Remark~\ref{rem2.3}, the diffusion exponents converge to zero for energy
intervals $I$ approaching the bottom of the spectrum.

%%% Local Variables: 
%%% mode: latex
%%% TeX-master: "newdyn"
%%% End: 

\section{Proof of the Main results}
%The first two lemmas are the key results of this section. Together
%with Lemma~\ref{heisenberg} and Lemma~\ref{combes-montcho},
%respectively, they imply Theorem~\ref{maintheorem} on dynamical
%localization and Theorem~\ref{maintheorem2} on the absence of
%diffusion.

Lemma \ref{heisenberg} guarantees that the Cesaro mean $\la\la
|X|^2\ra\ra_{T,\vp}$ of the second
moment of position is well-defined for all $T>1$ and all $\varphi$ as
in Theorem~\ref{maintheorem}. Lemma~\ref{combes-montcho} 
exhibits the relation between the asymptotic
behavior in $T$ of $\la\la
|X|^2\ra\ra_{T,\vp}$, for $\vp$ localized in energy in a compact set
$I$, and the behavior of the resolvents $(H-E-\imath\eps)^{-1}$ for
energies $E+\imath\eps$ approaching the real axis. In particular, it
shows that the main contribution to the second moment of position is
due to energies $E$ in $I$. The proof of Theorem~\ref{maintheorem} is
then completed with the help of Lemma~\ref{msaLemma}. To show
Theorem~\ref{maindiscrete}, we refer in addtion to
Remark~\ref{remdynloc}.i). On the other hand, 
Theorem~\ref{maintheorem2} follows from the first two lemmas and
Lemma~\ref{msaLemma2}.

The first lemma extends the results of Radin and Simon
\cite{radinsimon} to a larger class of operators $H$, although the
set $S$ we consider is slightly different.

%%%%%%%%%%%%%%%%%%%%%%% Lemma 4.1 %%%%%%%%%%%%%%%%%%%%%%%%%%%%%%%%%
\begin{lem}\label{heisenberg}
  Let $H^{(\omega)}$ be a random operator satisfying Assumptions 
  \ref{ass}.i)--iv)
  and let
  \begin{eqnarray}
    S := \{ \vp\in {\cal H},\  \pa \, 
    \vert X \vert \vp \pa <\infty,\ \pa
    H^{(\omega)}\vp\pa <\infty \} \ .\label{heisenberg1} 
  \end{eqnarray}
  \noindent
  Then for $\bbP$-a.e. $\omega$, $e^{-\imath t H^{(\omega)}}$ maps $S$
  into $S$, and there exists a finite constant $c^{(\omega)}$ such that for all
  $t\in \bbR$, $d=1,\dots ,\, j$, $\vp\in S$ and $\psi\in {\cal H}$, one has
  \begin{equation}
    \la\psi,\, [\e^{-\imath t H^{(\omega)}}, X_j]\vp\ra  = -\imath\int_0^t
    \la\psi,\, \e^{-\imath (t-s)H^{(\omega)}} [H^{(\omega)},X_j]
    \e^{-\imath s H^{(\omega)}} \vp\ra\, ds \ ,
    \label{heisenberg2}
  \end{equation}
  and
  \begin{equation}
    \pa\, \vert X\vert \e^{-\imath t H^{(\omega)}} \vp \pa  \leq  
    c^{(\omega)} \vert  t \vert\, \big(\pa \vp \pa + \pa H^{(\omega)}\vp
      \pa \big) + \pa \, \vert X \vert \vp \pa \label{heisenberg3}\ .
  \end{equation}
\end{lem}
%%%%%%%%%%%%%%%%%%%%% end of lemma 4.1 %%%%%%%%%%%%%%%%%%%%%%%%%%

%%%%%%%%%%%%%%%%%%%%%%%%%%%%%%%%%%%%%%%%%%%%%%%%%%%%%%%%%%%%%%%%%
\begin{rem} 
  Strictly speaking, the set $S$ depends on $\omega$, but due to
  Assumption \ref{ass}.i), $S$ is $\bbP$-almost surely constant.
\end{rem}

%%%%%%%%%%%%%%%%%%%  Proof of lemma 4.1   %%%%%%%%%%%%%%%%%%%%%%%%
\begin{proof} 
  Since there is no possible confusion, we drop the superscript $\omega$.
  Fix $j\in\{1,\dots ,\, d\}$ and let
  $$
  F_\eps = (H+\imath)\frac{1}{\imath + \eps
    X_j}(H+\imath)^{-1}\ .
  $$
  By Assumption \ref{ass}.ii) and since ${\cal C}$ is a core for
  $H$, the domain of $F_\eps$ contains the dense subset ${\cal
    B} := (H+\imath){\cal C}$. Now on $B$, one can write
  \begin{eqnarray}\label{heisenberg4}
    F_\eps := -\frac{\eps}{\imath + \eps X_j}[H,
    X_j](H+\imath)^{-1}F_\eps+\frac{1}{\imath + \eps X_j}\ .
  \end{eqnarray}  
  Thus for $\eps<\eps_1$ small enough, one has
  \begin{eqnarray*}
    F_\eps=\Big( 1+\frac{\eps}{\imath + \eps
      X_j}[H,X_j](H+\imath)^{-1}\Big)^{-1}
    \frac{1}{\imath + \eps X_j}\ ,
  \end{eqnarray*}
  which implies that $F_\eps$ is uniformly bounded in $\eps$.
  From \eqref{heisenberg4}
  one obtains
  \begin{equation}
    (\imath + \eps X_j)F_\eps  =   1-\eps [H,X_j](H+\imath)^{-1}
      F_\eps \nonumber\ ,
  \end{equation}
  which is an operator uniformly bounded in $\eps$ for all $\eps < \eps_2$
  small enough.  
  Hence, the operator
  \begin{eqnarray}\label{heisenberg5}
    (H+\imath )\frac{X_j}{\imath+\eps X_j}(H+\imath)^{-1} & = & [H,X_j
    ](H+\imath )^{-1} F_\eps + X_j F_\eps 
  \end{eqnarray}
  is bounded. Now one has
  \begin{equation}\label{heisenberg6}
    \frac{X_j}{\imath+\eps X_j} \e^{-\imath t H}
     =   \e^{-\imath t H}\Big( \e^{\imath
      t H}\frac{X_j}{\imath+\eps X_j}
    \e^{-\imath t H} - \frac{X_j}{\imath+\eps X_j}
    \Big) + \e^{-\imath t H} \frac{X_j}{\imath+\eps X_j}\ .
  \end{equation}
  From \eqref{heisenberg5}, since $X_j / (\imath + \eps X_j)$ maps
  ${\cal D}(H)$ into ${\cal D}(H)$, one obtains for the term in
  parenthesis in the right hand side of \eqref{heisenberg6}, in the
  weak sense
  \begin{equation}\label{heisenberg7} 
    \e^{\imath  t H}\frac{X_j}{\imath+\eps X_j} \e^{-\imath t H} -
    \frac{X_j}{\imath+\eps X_j} = \imath\int_0^t \e^{\imath s
      H} [H,\frac{X_j}{\imath + \eps X_j}]\e^{-\imath s H}  \, ds\ .
  \end{equation}
  Furthermore, by writing 
  \begin{equation*}
    [H,\frac{X_j}{\imath + \eps X_j}] =  [H, X_j] \frac{1}{\imath + \eps
      X_j} + X_j   [H,\frac{1}{\imath + \eps X_j}]
  \end{equation*}
  and
  \begin{eqnarray*}
    \lefteqn{X_j   [H,\frac{1}{\imath + \eps X_j}]} \nonumber \\
    & = & X_j \left(
      \frac{1}{\imath + \eps X_j} (\imath + \eps X_j) H \frac{1}{\imath +
        \eps X_j} -  \frac{1}{\imath + \eps X_j} H (\imath + \eps
      X_j) \frac{1}{\imath + \eps X_j} \right)\nonumber \\
    & = & X_j  \frac{1}{\imath + \eps X_j} [\eps X_j, H]   \frac{1}{\imath
      + \eps X_j} \ ,
  \end{eqnarray*}
  one obtains, together with  \eqref{heisenberg6} and
  \eqref{heisenberg7} 
  \begin{eqnarray*}
    \!\!\!\!\!\!\!\frac{X_j}{\imath+\eps X_j} \e^{-\imath t H}  &\!\!\!
    = &\!\!\! \imath\,\e^{-\imath t H}\int_0^t \!\!\e^{\imath s H}
    [H,X_j](H+\imath)^{-1}(H+\imath)\frac{1} {\imath+\eps X_j}
    \e^{-\imath s H}\, ds \\ 
    & \!\!\!&\!\!\!\!\!\!  +\imath\int_0^t \!\!e^{\imath s H} X_j
    \frac{\eps}{\imath+\eps 
    X_j}[X_j, H]
    (H+\imath)^{-1}(H+\imath) \frac{1}{\imath+\eps X_j}e^{-\imath s H} ds \\
    & & \!\!\!\!\!\!+\, \e^{-\imath t H} \frac{X_j}{\imath+\eps X_j}\ .
  \end{eqnarray*}
  Then for all $\vp\in{\cal D}(H)$,
  \begin{eqnarray*}
    \pa \frac{X_j}{\imath + \eps X_j }\e^{-\imath t H}\vp \pa & \leq & |t|\, 
    \pa [H, X_j ](H+\imath)^{-1}\pa\, \pa F_\eps\pa \, \pa (H+\imath )
    \vp\pa \nonumber \\
    & & + |t|\, \pa \frac{\eps X_j}{\imath + \eps X_j} \pa \, 
    \pa [H, X_j ](H+\imath)^{-1}\pa\, \pa F_\eps\pa \, \pa (H+\imath )
    \vp\pa \\
    & & + \pa \frac{X_j}{\imath + \eps X_j}\vp \pa   \\
    & \leq & c |t|\, \big(\pa H\vp \pa + \pa \vp \pa \big) 
    + \pa \frac{X_j}{\imath+\eps X_j}  \vp \pa \ .
  \end{eqnarray*}
  The constant $c$ is independent of $\eps$. Taking the limit $\eps\to
  0$, one gets for all $\vp\in S$ 
  \begin{eqnarray*}
    \pa X_j \e^{-\imath t H}\vp \pa \leq c |t|\, \big(\pa H\vp \pa +
    \pa \vp \pa\big) 
    +
    \pa X_j \vp \pa\ .
  \end{eqnarray*}
  This proves \eqref{heisenberg3}. Now, since we know that $\e^{-\imath t H}$
  maps
  $S$ into $S$, one has for all $\vp$ and $\psi$ in $S$:
  \begin{eqnarray}
    \la \e^{\imath t H} X_j \e^{-\imath t H}\vp,
    \, \psi\ra & = & \la X_j \vp, \, \psi\ra + \imath \int_0^t \la
    \e^{\imath s H} [H,X_j] \e^{-\imath s H}\vp, \, \psi \ra \, ds \ .
    \label{heisenberg8}
  \end{eqnarray}
  Since the integrand in \eqref{heisenberg8} is uniformly bounded in
  $s\in[0,t]$ and since $S$ is dense in ${\cal H}$, one can extend 
  \eqref{heisenberg8} to all $\psi \in {\cal H}$, 
  which gives \eqref{heisenberg2}.
\end{proof}
%%%%%%%%%%%%%%%%%%%% end of proof of lemma 4.1 %%%%%%%%%%%%%%%%%

The following lemma  is an easy generalization of a result
of Montcho \cite{M} stated for operators $H=-\Delta + V$, where $V$ is
bounded below. See also \cite{holdenmartinelli} for related results in
the case of random potentials which are piecewise constant.

%%%%%%%%%%%%%%%%%%%%%% Lemma 4.2 %%%%%%%%%%%%%%%%%%%%%%%%%%%%%%%% 
\begin{lem}\label{combes-montcho}
  Let $H$ be a random operator satisfying Assumptions
  \ref{ass}.i)--iv).  Let $f\in C_0^\infty(\bbR)$ be a non-negative
  function and 
  $I\supset {\rm supp\ }f$ any compact interval such that
  $\delta := {\rm dist}(\partial I, {\rm supp\ }f)>0$; then for all
  $\vp\in{\cal D}(H)\cap{\cal D}(\vert X \vert )$ there exist
  constants $c_1$,$c_2$ and $c_3$ depending only on $\vp$, $\delta$
  and $\pa f \pa_{L^\infty(\bbR)}$ such that for all $\eps := 1/T
  >0 $ we have, for $\bbP$ a.e. $\omega$:
  \begin{multline}
    \frac{1}{T}  \int_0^T \pa\,\vert X\vert \e^{-\imath t H^{(\omega)}}
    f(H^{(\omega)} )
    \vp \pa^2 dt \\
    \leq c_1 + c_2 b^2(\omega) + c_3 \eps \int_I \pa\, \vert X \vert
    (H^{(\omega)} -E+\imath
    \frac{\eps}{2})^{-1} \vp \pa^2
    dE\ \label{cm0}\ .
  \end{multline}
  The random variable $b(\omega)$ was defined in 
  Assumption~\ref{ass}.iii). 
\end{lem}
%%%%%%%%%%%%%%%%%%%%   end of lemma 4.2  %%%%%%%%%%%%%%%%%%%%%%%%

%%%%%%%%%%%%%%%%%%     Proof of lemma 4.2   %%%%%%%%%%%%%%%%%%%%%
\begin{proof}
  We will drop the label $\omega$ in the proof. Since $H$ is
  self-adjoint and $\vert X \vert$ is closed, we have for $\eps=1/T$
  \cite[Lemma~1 on p.\ 412]{reedsimon3}
  \begin{eqnarray}
    \frac{1}{T} \int_0^T \pa \,\vert X\vert \e^{-\imath t H} f(H)
    \vp \pa^2 dt & \leq & \e\, \eps \int_\bbR \pa\, \vert X \vert 
    \e^{-\eps \frac{t}{2}}\Theta (t) 
    \e^{-\imath t H} f(H)
    \vp \pa^2 dt \nonumber  \\
    & =  & \frac{ \e\, \eps}{2 \pi} \int_\bbR \pa \, \vert  X \vert R_\eps(E)
    f(H)
    \vp\pa^2 dE \label{peter} \ ,
  \end{eqnarray}
  where $R_\eps(E)=(H-E+\imath\eps /2)^{-1}$ and $\Theta(t)$ is the
  Heaviside function. In order to bound the right-hand side of
  \eqref{peter}, we split the range of integration over $E$ into two
  parts and fix $j\in\{1,\dots ,\, d \}$
  \begin{align}
    \int_{\bbR\setminus I}\pa X_j R_\eps(E) & f(H)\vp \pa^2 dE
    \nonumber \\ 
    & \leq  4\left( 
    \int_{\bbR\setminus I} \pa R_\eps(E)f(H)  X_j
    \vp \pa^2 dE \right.\label{cm1} \\
    & \qquad + \int_{\bbR\setminus I}\pa [X_j,f^{\frac{1}{2}}(H)]
    f^{\frac{1}{2}}(H) 
    R_\eps(E) \vp\pa^2 dE \label{cm2} \\
    & \qquad + \int_{\bbR\setminus I}\pa
    f^{\frac{1}{2}}(H)R_\eps(E) [H, X_j] 
    R_\eps(E)f^{\frac{1}{2}}(H)\vp \pa^2 dE \label{cm3} \\
    & \qquad + \left. \int_{\bbR\setminus I}
    \pa f^{\frac{1}{4}}(H)R_\eps(E)
    f^{\frac{1}{4}}(H)[X_j, f^{\frac{1}{2}}(H)]\vp\pa^2 dE \right)\ .
    \label{cm4}
  \end{align} 
  Since ${\rm dist}(\partial I, {\rm supp\ }f)>0$, one can bound the
  term \eqref{cm1} from above by
  \begin{eqnarray}
    \int_{\bbR\setminus I} \pa R_\eps(E)f(H) \pa^2 \pa X_j \vp \pa^2 dE 
    & \leq & \frac{2 \pa f \pa_\infty^2}{\delta} \pa X_j \vp \pa^2
    \label{cm5} \ .
  \end{eqnarray}
  We now treat the term \eqref{cm2}. Let $\tilde{\vp}_E \equiv
  R_\eps(E) \vp$; then obviously, since $\vp$ is in ${\cal D}(H)$,
  $\pa \tilde{\vp}_E \pa$ and $\pa H \tilde{\vp}_E \pa$ are finite and
  from Lemma~\ref{heisenberg} one has, by using the equality $R_\eps(E) =
  -\imath \int_0^\infty \e^{-\eps t/2} \e^{\imath t (H-E)} dt$
  \begin{eqnarray}
    \pa\, \vert X \vert  \tilde{\vp}_E \pa & \leq & 
    \int_0^{+\infty} 
    \e^{-\frac{\eps  t}{2}}\pa \, \vert X \vert \e^{\imath t (H-E)}
    \vp \pa dt  \nonumber \\
    & \leq & c  \int_0^{+\infty} 
    \e^{-\frac{\eps  t}{2}} \Big(t (\pa\vp\pa + \pa H\vp\pa )+ \pa \,
    \vert
    X\vert \vp \pa) \Big) \nonumber \\
    & < & \infty\ .
  \end{eqnarray}
  \begin{sloppypar}
    \noindent Therefore $\tilde{\vp}_E$ is in the set $S$ defined in
    Lemma~\ref{heisenberg}. By writing $f^{1/2}(H)= \int_\bbR
    \widehat{f^{1/2}
      }(t)\e^{\imath t H} dt$,
                                %in the strong sense, 
    one can prove with the same arguments as above that
    $f^{1/2}(H)R_\eps(E)\vp\in S$. Then we get from Lemma~\ref{heisenberg}
  \end{sloppypar}
  \begin{eqnarray}
    \lefteqn{\pa[f^\frac{1}{2}(H),
      X_j]f^\frac{1}{2}R_\eps(E)\vp\pa}\nonumber \\
    & \leq & \sup_{\xi\in{\cal H},
      \pa\xi\pa=1}\int_\bbR \int_0^t
    \vert \widehat{f^\frac{1}{2}}(t) \vert \, 
    \vert \la\xi , \, \e^{-\imath (s-t) H}[H,X_j]
    \e^{\imath s H}f^\frac{1}{2}(H)R_\eps(E)\vp \ra\vert ds 
    \, dt  \nonumber \\
    & \leq &  \int_\bbR t \vert
    \widehat{f^\frac{1}{2}}(t)\vert 
    \, \pa [H, X_j](H+\imath)^{-1}\pa\, \pa (H+\imath)
    f^\frac{1}{4}(H)\pa\, \pa R_\eps(E) f^\frac{1}{4}(H)\vp\pa dt \ ,
    \nonumber 
  \end{eqnarray}
  and, using Assumption~\ref{ass}.iii), \eqref{cm2} is bounded
    above by
  \begin{multline}
    \int_{\bbR\setminus I}\pa[f^\frac{1}{2}(H),
      X_j]f^\frac{1}{2}R_\eps(E)\vp\pa^2 dE \\
     \leq  b^2 (\max_{E\in I} \vert E \vert + 1)^2
    \Big(\int_\bbR  t \vert \widehat{f^\frac{1}{2}} (t)\vert dt\Big)^2
    \ \frac{2\pa f \pa_\infty}{\delta} \ . \label{cm6} 
  \end{multline}
  To bound the term \eqref{cm3} one writes
  \begin{eqnarray}
    \lefteqn{\int_{\bbR\setminus I} \pa 
      f^\frac{1}{2}(H)R_\eps(E) [H, X_j]R_\eps(E)
      f^\frac{1}{2}(H)\vp\pa^2 dE } \nonumber \\
    & \leq & \int_{\bbR\setminus I} \Big( \pa f^\frac{1}{2}(H)R_\eps(E)\pa^2
    \pa[H, X_j](H+\imath)^{-1}\pa^2 \nonumber \\
    &  & \qquad\times \ \pa (H+\imath)f^\frac{1}{4}(H)\pa^2
    \pa f^\frac{1}{4}(H)R_\eps(E)\pa^2\pa\vp\pa^2 dE \Big) \nonumber \\
    & \leq & \frac{2 b^2}{3 \delta^3} \pa f\pa_\infty^2 (\max_{E\in
      I}\vert E\vert +1)^2 \pa\vp\pa^2 \ . \label{cm7}
  \end{eqnarray}
  Finally, for \eqref{cm4} one has
  \begin{eqnarray}
    \lefteqn{\int_{\bbR\setminus I}\pa 
      f^\frac{1}{4}(H)R_\eps(E)f^\frac{1}{4}(H)
      [X_j, f^\frac{1}{2}(H)]\vp\pa^2 dE } \nonumber \\
    & \leq & \Big( \int_{\bbR\setminus I} f^\frac{1}{4}(H)R_\eps(E)\pa^2
    dE\Big) \Big( \sup_{\xi\in {\cal H}, \pa\xi\pa=1} \vert \la
    f^\frac{1}{4}(H)\xi,\, [f^\frac{1}{2}(H), X_j]\pa\ra\vert \Big)^2
    \nonumber \\
    & \leq & \frac{2 \pa f\pa_\infty^\frac{1}{2}}{\delta} 
    \sup_{\xi\in {\cal H}} \Big( \vert \int_\bbR \widehat{f^\frac{1}{2}}(t) 
    \la f^\frac{1}{4}(H)\xi , \, 
    [\e^{\imath t H}, X_j]\vp \ra dt\vert\Big)^2 
    \nonumber \\
    & \leq & \frac{2 \pa f\pa_\infty^\frac{1}{2}}{\delta} 
    \Big( \sup_{\xi\in {\cal H}}\int_\bbR 
    \vert \widehat{f^\frac{1}{2}}(t) \vert 
    \int_0^t \vert\la f^\frac{1}{4}(H)
    \xi, \, \e^{-\imath (s-t) H}[H, X_j] 
    \e^{\imath s H}\vp\ra\vert ds\, 
    dt \Big)^2 \nonumber \\
    & \leq & \frac{2 b^2 \pa f\pa_\infty}
    {\delta} \max_{E\in I}(\vert
    E\vert+1)^2 \pa\vp\pa^2\Big(\int_{\bbR}
    t \vert \widehat{f^\frac{1}{2}}(t)\vert \Big)^2 \ .\label{cm8}
  \end{eqnarray}
  Inequalities \eqref{cm5}-\eqref{cm8} imply 
  that there exists $c<\infty$ independent of $\omega$ such that 
  \begin{eqnarray}
    \int_{\bbR\setminus I} \pa\, \vert X\vert \e^{-\imath t
    H}f(H)\vp\pa^2 dE  
    & \leq & c\, b^2\ . \label{cm9}
  \end{eqnarray}
  As to the remaining contribution to \eqref{peter}, we write
  \begin{multline}
    \eps\int_I \pa X_j R_\eps (E) f(H)\vp\pa^2 dE  \\
     \leq  2 \eps \int_I \pa [X_j, f(H)] R_\eps(E)\vp\pa^2 dE
    + 2 \eps \int_I \pa X_j R_\eps(E) \vp\pa^2 dE \ .\label{cm10}
  \end{multline}
  The first term in \eqref{cm10} is bounded from above by
  \begin{multline}
    4\eps\int_I\pa [X_j, f^\frac{1}{2}(H)]
    f^\frac{1}{2}(H)R_\eps(E)\vp\pa^2dE \\ + 4\eps\int_I\pa
    f^\frac{1}{2}(H)
    [X_j, f^\frac{1}{2}(H)]R_\eps(E)\vp\pa^2dE  \\
    \leq 4\eps\int_I\int_\bbR\vert
    \widehat{f^\frac{1}{2}} (t)\vert\,\pa [X_j,\e^{\imath t
      H}]f^\frac{1}{2}(H) R_\eps(E)\vp\pa^2 dt\, dE  \\
    + 4\eps\int_I \int_\bbR \vert \widehat{f^\frac{1}{2}} (t)
    \vert\,\pa f^\frac{1}{2}(H)[X_j,\e^{\imath t H}] R_\eps(E)\vp\pa^2
    dt\, dE \ .\label{cm11}
  \end{multline}
  According to the fact that $R_\eps(E)\vp$ and 
  $f^\frac{1}{2}(H) R_\eps(E)\vp$ are both in the set $S$ defined 
  in \eqref{heisenberg1}, one can prove as above with the
  help of Lemma~\ref{heisenberg} that each of the two terms in \eqref{cm11}
  are bounded  above by
  \begin{multline}
    \Big(\int_\bbR t\vert \widehat{f^\frac{1}{2}}(t)
      \vert dt\Big)^2 \pa [H,X_j] (H+\imath)^{-1} \pa^2 
    \; \pa (H+\imath)f^\frac{1}{2}(H)\pa^2 
    \eps \int_I \pa R_\eps(E)\vp\pa^2 dE  \\
     \leq   \pi\Big(\int_\bbR t\vert \widehat{f^\frac{1}{2}}(t)
    \vert dt\Big)^2 b^2 (\max_{E\in I} \vert E\vert +1)^2
      \|f\|_\infty\ . \nonumber
  \end{multline}
  This inequality together with \eqref{cm9} and \eqref{cm10} implies 
  \eqref{cm0}.
\end{proof}
%%%%%%%%%%%%%%%%%%% end of the proof of lemma 4.4  %%%%%%%%%%%%%%%%

\begin{rem}
  Although stated for random operators, Lemmas~\ref{heisenberg} and
  \ref{combes-montcho} are
  purely deterministic, in the sense that their statements remain true
  if we consider a non-random $H$ satisfying 
  Assumptions \ref{ass}.i)--iv). 
\end{rem}

The following two lemmas represent the key results of this section.

%%%%%%%%%%%%%%%%%%%%%%%%%%%%%  Lemma 4.3 %%%%%%%%%%%%%%%%%%%%%%%
\begin{lem}\label{msaLemma}
Let $H$ satisfy Assumptions \ref{ass}.i), v) and the Multi-Scale
Assumption (M1). Then there exists a constant $c_4 <\infty$ such that
for all $\varepsilon < 1$,
\begin{equation}
\label{msaLemmaBound}
\bbE\left\{\int_I \varepsilon \pa\, \vert X
\vert (H^{(\omega)}-E-\imath\varepsilon )^{-1} \vp \pa^2 d E \right\}
\le  c_4 \left( \log1/\varepsilon \right)^{\alpha(d+2)/\nu}
\end{equation}
for all normalized $\varphi\in\calH $ with compact support $\supp
\varphi $. The constant $c_4$ depends on $d$, the size of
$\supp \varphi$ and on the constants $\alpha$, $p$ and $\nu$ of the Multi-Scale
Assumption (M1). 
\end{lem}
\begin{proof}
Without loss of generality we assume that $\supp\varphi\subseteq \{ x
\ \vert\ {\pa}x\pa_\infty <L_0 \}$ and $\|\varphi \| \leq 1$. The
proof is done in three steps.  

First we consider $\omega\in \Omega_k
(q,q') $ with $ \pa q-q'\pa > 2 L_k$, where
  \begin{multline}
    \Omega_k (q,q') := \{ \omega \ \vert\  H^{(\omega)} \text{ is }
    (\rho,E,L_k,q'')-\text{regular for all } E\in I\\
    \text{ and either }
    q''=q \text{ or } q''=q'\}\, .\nonumber
  \end{multline}
  Let us introduce the following notation
$$
R^{(\omega)}(E+\imath\varepsilon) := (H^{(\omega)}-E-\imath \varepsilon)^{-1} .
$$
If $ H^{(\omega)}$ is $(\rho,E,L_k,q')$-regular, we apply 
  (\ref{resEqn}) to get
  \begin{equation}
    \label{qpregBound}
    \pa \one_{q'}  R^{(\omega)}(E+\imath\varepsilon)\one_q
    \pa^{2} \leq \rho(L_k) \pa  R^{(\omega)}(E+\imath\varepsilon)\one_q \pa^{2}. 
  \end{equation}
In case $ H^{(\omega)}$ is $(\rho,E,L_k,q)$-regular, we apply the
adjoint of (\ref{resEqn}) to obtain the bound (\ref{qpregBound})
with $q$ replaced by $q'$ and $\varepsilon$ replaced by
$-\varepsilon$ on the right-hand side of (\ref{qpregBound}).  In any
case one has for all $\omega\in\Omega_k(q,q')$, with $\pa q-q'\pa >
2 L_k $, the estimate
\begin{equation}\label{anyCase}
\pa \one_{q'} R^{(\omega)}(E+\imath\varepsilon) \one_q \pa^{2}\leq
\rho(L_k) \left( \pa R^{(\omega)}(E+\imath\varepsilon) \one_q
\pa^{2} + \pa R^{(\omega)}(E+\imath\varepsilon) \one_{q'} \pa^{2}
\right) \, .
\end{equation}
Second we derive the upper bound
\begin{equation}\label{secondBound}
\bbE\left\{ \int_I \pa \one_{q'}
R^{(\omega)}(E+\imath\varepsilon)\vp \pa^2 d E \right\} \le (2 L_0
+1)^{2d} \left(2\rho(L_k)\varepsilon^{-2} |I| + \pi L_k^{-p}
\varepsilon^{-1}\right)
\end{equation} 
for all $q'\in\bbZ^{d} $ with $L_0 + 2 L_k < \pa
q' \pa_\infty \le L_0 + 2 L_{k+1}$.  To this end we note that 
\begin{equation}\label{suppSum}
    \pa\one_{q'} R^{(\omega)}(E+\imath\varepsilon)\vp \pa^2 \le\!  (2
    L_0+1)^{d}\!\!\!\!\!\sum_{q\in\Lambda_{L_0}(0)\cap\bbZ^d}
    \!\!\!\!\!\pa\one_{q'}
    R^{(\omega)}(E+\imath\varepsilon)\one_{q}\pa^2\, \pa\one_{q}\vp
    \pa^2 \,,
\end{equation} 
where $\Lambda_{L_0}(0)$ is defined as in Section 3. Then,
\begin{eqnarray}\label{eq:star}
\lefteqn{\int_{\Omega} \left(\int_I \| \one_{q'}
R^{(\omega)}(E+\imath\varepsilon)\vp\|^2 dE \right) d\bbP(\omega)} & &
\nonumber \\
& \leq & (2L_0+1)^d \sum_{q\in \Lambda_{L_0}(0)\cap\bbZ^d}
\int_{\Omega} \left(\int_I \| \one_{q'}
R^{(\omega)}(E+\imath\varepsilon)\one_{q}\vp\|^2 dE \right)
d\bbP(\omega) \nonumber \\ 
& \leq & (2L_0+1)^d \sum_{q\in
\Lambda_{L_0}(0)\cap\bbZ^d} \Bigg\{ \int_{\Omega_k(q,q`)} \bigg(\int_I
\rho(L_k) \Big( \| R^{(\omega)}(E+\imath\varepsilon)\one_{q}\|^2 
\nonumber\\
& &  \hspace*{2cm} + \|
R^{(\omega)}(E+\imath\varepsilon)\one_{q'}\|^2\Big) dE \bigg) 
d\bbP(\omega) \nonumber\\ 
& & \hspace*{2cm} + \int_{\Omega\setminus\Omega_k(q,q')}
\left(\int_I \| \one_{q'}
R^{(\omega)}(E+\imath\varepsilon)\one_{q}\vp\|^2 dE \right)
d\bbP(\omega) \Bigg\}
\end{eqnarray}
where we have used \eqref{anyCase} in the integral over $\omega\in
\Omega_{k}(q,q')$. For these $\omega$ we proceed by bounding the
resolvents according to $\|R^{(\omega)}(E+\imath\varepsilon)\|\leq
\eps^{-1}$. This yields the term $2\rho(L_k) \eps^{-2} |I|$ in
\eqref{secondBound}. For $\omega\notin\Omega_{k}(q,q')$ we use  
\begin{equation}
\label{residuum}
\int_I \, \pa\, R^{(\omega)}(E+\imath\varepsilon) \psi \pa^2 d
E \le \frac{\pi}{\varepsilon}\,,
\end{equation}
which is valid for all $\psi \in \calH $ with $\pa\psi\pa=1 $, and the
Multi-Scale Assumption (M1), viz.\ 
$\bbP(\Omega\setminus\Omega_k(q,q')) \leq L_k^{-p}$, giving the second
term in \eqref{secondBound}.

Finally we derive \eqref{msaLemmaBound} as follows. Given
$\eps$, there exists $k=k(\eps)\in\bbN$ such that 
\begin{equation} \label{keps}
\e^{L_{k-1}^{\,\nu}}\leq \eps^{-1} < \e^{L_{k}^{\,\nu}}\,.
\end{equation}
Thus we get, for some constants $c_0$ and $c_4$ 
\begin{align*}
  \bbE \Bigg\{ \int_I  &  \varepsilon \pa\, \vert X \vert
    R^{(\omega)}(E+\imath\varepsilon) \vp \pa^2 d E \Bigg\}
  \nonumber \\ 
  \le\; &\eps \bbE\left\{ \int_I
    \|\chi_{\Lambda_{L_0+2L_k}(0)} |X| R^{(\omega)}(E+\imath\varepsilon)
    \vp \|^2 dE \right\}  \nonumber\\ 
  &  + \eps
  \sum_{j=k}^{\infty}\ \sum_{L_0+2L_j < \|q'\|_\infty\leq
    L_0 + 2 L_{j+1}}\! \bbE\left\{ \int_I \| \one_{q'} |X|
    R^{(\omega)}(E+\imath\varepsilon) \vp \|^2 dE \right\}
  \displaybreak[0]\\
  \le\; &  c_0 \left(L_k^{d+2} + \sum_{j=k}^{\infty} 
    L_{j+1}^{d+2} \left(\rho(L_j)\eps^{-1} |I| + L_j^{-p}\right)  \right)
  \displaybreak[0]\\
  \le\; & c_0\left( (\log 1/\eps)^{\alpha (d+2)/\nu} + 
    \sum_{j=k}^{\infty} 
    L_{j}^{\alpha(d+2)} \left(\e^{-2L_j^{\,\nu}}\e^{L_j^{\,\nu}} |I| 
      + L_j^{-p}\right) \right) \\
  \le\; & c_4 (\log 1/\eps)^{\alpha(d+2)/\nu} \,.
\end{align*}
\end{proof}
\begin{rem}\label{remdynloc}
\begin{enumerate}
\item[i)]
If ${\cal H}=\ell^2(\bbZ^d)$, one can also use \eqref{residuum} in the
integral over $\omega\in\Omega_{k}(q,q')$ in \eqref{eq:star} and
derive the bound $2\pi\rho(L_k)\eps^{-1}$ instead of
$2\rho(L_k)\eps^{-2}|I|$ for the first term in \eqref{secondBound}. 
This implies that \eqref{msaLemmaBound} is bounded
uniformly in $\eps$. Thus, we obtain the dynamical localization
property \eqref{13bis}, as claimed in Theorem~\ref{maindiscrete}.
\item[ii)]
If the decay of the function $\rho$ in the Multi-Scale Assumption (M1)
is only algebraic
\begin{equation}
  \rho(x) \le c(n) \, x^{-2n}\,,
\end{equation}
for some $n\in\bbN$ with $n>\alpha (d+2)$ and some constant
$c(n)<\infty$, then choosing $k(\eps)\in\bbN$ according to 
$L_{k-1}^{n} < \eps^{-1} \le L_{k}^{n}$ instead of \eqref{keps}, we bound 
\eqref{msaLemmaBound} by $c_{4}\eps^{-\alpha (d+2)/n}$. This implies 
$\sigma^+_{{\rm diff}}\bigl(f_{I'}(H)\varphi\bigr)
\leq\frac{\alpha(d+2)}{n}$, as claimed in Remark~\ref{rem2.3}.
\end{enumerate}
\end{rem}           
%%%%%%%%%%%%%%%%%%%%%  end of lemma 4.3 %%%%%%%%%%%%%%%%%%%%%%%%

%%%%%%%%%%%%%%%%%%%%%% lemma 4.4 %%%%%%%%%%%%%%%%%%%%%%%%%%%%%
\begin{lem} \label{msaLemma2}
  Let $H$ satisfy Assumptions ~\ref{ass}.i), v) and (M2), then for
  $\varphi\in{\cal H}$, normalized and compactly supported, there
  exist $c_2<\infty$ and $\upsilon>0$ such that for all $0<\eps<1$
  $$
  \bbE\left\{\eps^2\int_I \pa\, |X| 
    (H^{(\omega)}-E-\imath\eps)^{-1}\varphi\pa^2 d E  \right\} <
  \eps^{\upsilon}\ .
  $$ 
\end{lem}
\begin{proof} We assume here, without loss of generality, that
  $\supp(\varphi)\subset\{x\  
  \vert\ \pa x \pa_\infty \leq L_0\}$; then we obtain, for $E\in I$,
  $\eps>0$ and $J_0\in\bbN$
  \begin{eqnarray*}
    \lefteqn{\pa \vert X \vert (H^{(\omega)} -E-\imath \eps)^{-1} \varphi
      \pa^2} \nonumber \\
    & \leq & (2 L_0+1)^2\sum_{k\geq J_0}\ \sum_{q\in\supp(\varphi)\cap\bbZ^d}\ 
    \sum_{q'\in\bbZ^d, L_0+2L_k <\pa q' \pa_\infty \leq L_0+2L_{k+1}}
    \!\!\!\!\!
    \nonumber \\
    & &  \quad\qquad\bigg\{ \pa\, \vert X \vert \unc_{q'} \pa_\infty^2 \pa
    \unc_{q'}(H^{(\omega)} -E-\imath 
    \eps)^{-1} \unc_q\pa^2 \pa \unc_q\varphi\pa^2\bigg\} \nonumber \\
    & & + \; d (L_{J_0} + L_0 +1)^2 \pa (H^{(\omega)} -E-\imath
    \eps)^{-1} \varphi \pa^2 \ . 
  \end{eqnarray*}
  For fixed $E$ and $q$  we define
  $$
  \Omega(E,q) := \{\omega\ \vert\ H^{(\omega)}\ {\rm is\ }(\rho,E,L_k,q){\rm
    -regular}  \}\ .
  $$
  Thus, according to Assumption (M2), we get 
  \begin{eqnarray}
    \lefteqn{\eps^2\int\left( \int_{I}  \pa \vert X \vert
        (H^{(\omega)}-E-\imath\eps)^{-1}\vp \pa^2 d E \right) d \bbP(\omega)}
    \nonumber \\ 
    &\!\! \leq & \!\!(2 L_0+1)^2 \sum_{k\geq J_0}\ \sum_{q\in\supp(\vp)\cap\bbZ^d}\
    \sum_{q',\, L_0+2L_k\leq \pa q'\pa_\infty \leq L_0+2L_{k+1}} \nonumber
    \\  
    &\!\!\! & \!\!\!\!\!\!\Bigg\{ \int_I \bigg(\int_{\Omega(E,q)}
    \!\!\!\!\frac{d (L_0+2L_{k+1}+1)^2} 
    {L_k^{m}} \eps^2 \pa (H^{(\omega)}-E-\imath\eps)^{-1} \pa^2
    d \bbP(\omega) \nonumber 
    \\
    &\!\!\! & \!\!\!\!\!\!+ \int_{\Omega^c(E,q)}
    \!\!\!\!\!\!\!d (L_0+2L_{k+1}+1)^2 \eps^2 \pa  
    \unc_{q'} (H^{(\omega)}-E-\imath\eps)^{-1} \pa^2 
    d\bbP(\omega)\bigg) dE \Bigg\} \label{aod1} \\ 
    &\!\!\! & \!\!\!\!\!\!+ d(L_0 + 2L_{J_0} + 1)^2 \bbE\left\{ \eps^2
      \int_I \pa 
      (H^{(\omega)}-E-\imath\eps)^{-1} \vp \pa^2 dE   \right\}\,.
    \label{aod2}
  \end{eqnarray}
  By using the inequality $\pa (H^{(\omega)}-E-\imath\eps)^{-1}\pa
  \leq \eps^{-1}$ in \eqref{aod1} and inequality \eqref{residuum} in
  \eqref{aod2}, we derive the upper bound
  $$
  \eps^2\bbE\left\{ \int_{I}  \pa \vert X \vert
    (H^{(\omega)}-E-\imath\eps)^{-1}\vp \pa^2 dE \right\} 
  \leq C (\eps L_{J_0}^2+\vert I\vert L_{J_0}^{-\beta})\ .
  $$
  Here $C$ is a constant depending only on $L_0$ and $d$. This
  gives the expected result for an appropriate dependence of $J_0$ on
  $\eps$: for $K\in\bbN$ such that $2\alpha^{-K}<1$ and if
  $L_J<\eps^{-1}\leq L_{J+1}$, we take $J_0=J-K$. We thus obtain
  $\upsilon=\min\{\beta\alpha^{-K-1},\, 1-2\alpha^{-K} \}$.
\end{proof}

%%% Local Variables: 
%%% mode: latex
%%% TeX-master: "newdyn"
%%% End: 

\appendix
\section{Appendix}

We present here the two lemmas necessary for the ``variable-energy''
multi-scale analysis needed in Example 3. For the rest of this section,
we fix $\alpha>1$, $\ell<\infty$, $N>4$ and $S$ even, $2<S<N-1$. For
$L=N\ell^\alpha$ we define for all $n\in\{ 0,1, \cdots ,N \}$
\begin{equation}
  \chi_n^L(x)= 
  \begin{cases}
    J_{\Lambda_{L/(4N)}(x),0} & n=0 \ ,\\
    J_{\Lambda_{nL/N}(x),\frac{(L/N)^{1/\alpha}}{4N}} & 1\leq n\leq
    N-1 \ ,\\ 
    J_{\Lambda_L(x),L/(4N)} & n=N \ ,\label{caracteristic}
  \end{cases}
\end{equation}
where, for $\delta> 0$, $J_{\Lambda ,\delta}$ is a smooth
characteristic function such that, for $\partial\Lambda$ being the
boundary of the box $\Lambda$,
\begin{equation}
  J_{\Lambda ,\delta}(y)=
  \begin{cases}
    1 & {\rm if}\ y\in\Lambda\ {\rm and}\
    \dist(y,\partial\Lambda)<\delta\ 
    ,\\
    0 & {\rm if}\ y\not\in\Lambda\ ,\nonumber
  \end{cases}
\end{equation}
and for $\delta=0$, $J_{\Lambda , \delta}$ is the characteristic
function of $\Lambda$. One also defines the frame $\gF_L(x,n)$ of the
box $\Lambda_{\frac{nL}{N}}(x)$ as the set of
$y\in\Lambda_{\frac{nL}{N}}(x) \cap(\frac{\ell}{4N}\bbZ)^d$ such that
$nL/N-2\ell/(4N)<\pa x-y\pa_\infty \leq nL/N-\ell /(4N)$. Thus $\supp
\nabla\chi_n^L(x)\subset\cup_{y\in\gF_L(x,n)} \Lambda_{\ell/(4N)}(y)$.
Note that Card($\gF$)$\leq c_{d,N} L^{(1-1/\alpha)(d-1)}$, $c_{d,N}$
being a constant depending only on $d$ and $N$. Now, for $w,m>0$ and
$n_i\in\{0,1,\dots,N\}$, $i\in\{1,\dots,S\}$, we define
$\gW_L(E,x,w,(n_i)_{i=1, \dots ,S})$ to be the property: $\forall
i=1,\dots ,S$,
$$
\sup_{\eps>0} \pa(H_{\Lambda_{\frac{n_iL}{N}(x)}}-E-\imath\eps)^{-1}
\pa \leq L^w\ ,
$$
and $\gH_L(E,x,m,(n_i)_{i=1, \dots ,S})$ as the property: $\forall
i=1,\dots ,S\ ,\ \forall y\in\gF_L(x,n_i)$,
\begin{equation}\label{equ.b}
\sup_{\eps>0} \pa [-\Delta, \chi_N^\ell(y)]
(H_{\Lambda_\ell(y)}-E-\imath\eps)^{-1} \chi_0^\ell(y) \pa \leq
\ell^{-m}\ . 
\end{equation}
Here $H_\Lambda$ refers to $H$ restricted to $\Lambda$ with
Dirichlet boundary conditions. We call a box $\Lambda_\ell(y)$ having
this last property an $(m,E)$-good box.
\begin{lem}\label{deterministiclemma} (Deterministic estimates).\\
  Given $x\in\bbZ^d$, $L<\infty$ and $E\in\bbR$, if
  $(S-\alpha)m>\alpha(S+1)w + S(d-1)(\alpha-1)$ and if there are $S$
  integers $(n_i)_{i=1,\dots ,S}$, $1\leq n_1<n_2<\dots <n_S\leq N-1$
  such that $\gW_L(E,x,w,(n_i))$ and $\gH_L(E,x,m,(n_i))$ hold, then
  \begin{eqnarray*}
    \sup_{\eps>0} \pa [-\Delta,\, \chi_N^L(x)] (H_{\Lambda_L(x)} -E 
    -\imath\eps)^{-1} \chi_0^L(x) \pa \leq L^{-m}\ .
  \end{eqnarray*} 
\end{lem}
The proof of the deterministic estimates is now well-known (see e.g.
\cite{vondreifusklein2} for discrete models and \cite{FLM} for
continuous models). For self-consistency, we give here the most
important steps of the proof of Lemma~\ref{deterministiclemma}.

\begin{proof} Let $W_{L,n,x}=[-\Delta,
  \chi_n^L(x)]$, ($n\neq 0$) and $R_\Lambda=(H_\Lambda-E-\imath\eps)^{-1}$.
  We first apply $S$ times the geometric resolvent equation to
  $W_{L,N,x}R_{\Lambda_L(x)}\chi_0^L(x)$ with the
  increasing sequence of boxes $\Lambda_{n_iL/N}(x)$, $i=1,\dots ,S$,
  \begin{eqnarray}
    \lefteqn{\!\!\!\!\!\!\!W_{L,N,x}R_{\Lambda_L(x)}\chi_0^L(x) }
    \nonumber \\ 
    & = &
    W_{L,N,x} R_{\Lambda_{\frac{n_S L}{N}}(x)} \chi^\partial_{n_S}(x)
    W_{L,n_S,x} 
    R_{\Lambda_{\frac{n_{S-1} L}{N}}(x)} \chi_{n_{S-1}}^\delta \ \ 
    \nonumber \\
    & &  \dots W_{L,n_2,x} R_{\Lambda_{\frac{n_{1} L}{N}}(x)}
    \chi_{n_1}^\partial(x) W_{L,n_1,x} 
    R_{\Lambda_{\frac{n_1 L}{N}}(x)} 
    \chi_0^L(x) \ , \ \label{GRE1}
  \end{eqnarray}
  where $\chi_{n_i}^\partial(x)$ is the characteristic function of the
  support of $\nabla\chi_{n_i}^L(x)$. Now, again with the help of
  the geometric resolvent equation, observe that for $j=1,\dots, S$
  \begin{eqnarray}
    \lefteqn{
      \pa W_{L,n_j,x} R_{\Lambda_{\frac{n_jL}{N}}(x)}
      \chi_{n_{j-1}}^\partial \pa
      } \nonumber  \\ 
    & & \leq \!\!\sum_{y\in \gF_L(x,n_{j-1})}\!\!\!\! \pa
    W_{L, n_j, x} R_{\Lambda_{\frac{n_j L}{N}}(x)} \chi_0^\ell(y) \pa
    \nonumber \\ 
    & & \leq  \!\!\sum_{y\in \gF_L(x,n_{j-1})}\!\!\!\! \pa 
    W_{L, n_j, x} R_{\Lambda_{\frac{n_j L}{N}}(x)} \chi_N^\ell(y)\pa\,
      \pa 
    W_{\ell, N, y} R_{\Lambda_\ell}(y) \chi_0^\ell(y) \pa\
      .\label{GRE2} 
  \end{eqnarray}
  \begin{sloppypar}
    In each term of this last sum, the first factor is estimated by
    using $\gW_L(E, x, w, (n_i))$, and the second factor by using
    $\gH_L(E,x,m,(n_i))$. Thus, \eqref{GRE1} together with \eqref{GRE2}
    gives
  \end{sloppypar}
  \begin{equation}
    \sup_{\eps>0} \pa W_{L, N, x} R_{\Lambda_L(x)} \chi_0^L \pa  \leq
    (\tilde{c}\, c_{d,N})^S \tilde{c} \big(L^{(1-\frac{1}{\alpha})(d-1)}
    L^w \ell^{-m} \big)^S L^w  \leq  L^{-m} \nonumber
  \end{equation}
  for $L$ large enough. Remark that here we have used the following
  inequality (see e.g. \cite[Appendix 1]{CH1} or \cite{FLM}), for all
  $L'$, $z$ and $n\in \{ 1,\dots ,N-1 \}$:
  $$
  \pa W_{L', n, z} R_{\Lambda_{L'}}(z) \chi_0^{L'} \pa \leq \tilde{c}
  \pa  
  R_{\Lambda_{L'}}(z)\pa\ .
  $$
\end{proof}
We now state the probabilistic part of this analysis. Since
\eqref{msaAss} requires a result which is uniform in energy, we must
do a ``variable-energy'' multi-scale analysis like in
\cite{vondreifusklein1}, adapted to correlated potentials. One of the
key estimates we use here is a correlated Wegner estimate proven in
\cite{CHM1}:
\begin{prop}\label{CHMestimate} 
(Combes-Hislop-Mourre correlated Wegner estimate).\\
We assume (A1) -- (A4) for the model \eqref{model2}. Then there exist
$I=[\alpha, \beta]$, a compact interval in the almost-sure spectrum of
$H^{(\omega)}$, $\ell_0 < \infty$ and $C_W>0$ such that for $E\in I$,
for any bounded open cubes $\Lambda_1$, $\Lambda_2$ with
$\dist(\Lambda_1, \Lambda_2)\geq \ell_0$ and for all $0<\eta<1$
$$
    \bbE\big(\Tr\left(\mathbf{E}_1[E-\eta, E+\eta]\right)
    \Tr\left(\mathbf{E}_2[E-\eta, E+\eta]\right)\big)\leq
    C_W\eta^2\vert\Lambda_1\vert \vert \Lambda_2\vert\ ,
$$  
where $\mathbf{E}_1$ and $\mathbf{E}_2$ are, respectively, the
spectral families associated to $H_{\Lambda_1}$ and $H_{\Lambda_2}$.
\end{prop}

We can then establish the following lemma:
\begin{lem}\label{probabilisticlemma} (Probabilistic estimates).\\
With the same notations as before, if there exist $L_0<\infty$ and
$p>0$ such that ${(\alpha-1)(d-1)(N-S)}/({\theta(N-S,
\alpha)-\alpha})< p< w-2d$, and for all $x_1$, $x_2$, $\pa x_1 - x_2
\pa_\infty > 2L_0$,
\begin{eqnarray*}
\bbP\Big\{ \forall E\in I,\, \exists (n_i)_{i=1,\dots S}\ {\rm s.  t.\
} \big( \gW_{L_0}(E, x_1, w, (n_i))\ {\rm and\ }
\gH_{L_0}(E,x_1,m,(n_i)) \ \big) \nonumber \\ \ \ {\rm or\ } \big(
\gW_{L_0}(E, x_2, w, (n_i))\ {\rm and\ } \gH_{L_0}(E,x_2,m,(n_i)) \
\big) \Big\} \geq 1-L_0^{-p}\ ,
\end{eqnarray*}
then for $L_1=N L_0^\alpha$ and for all $y_1$, $y_2$, $\pa y_1 - y_2
\pa_\infty > 2 L_1$, we have:
\begin{eqnarray}
\bbP\Big\{ \forall E\in I,\ \exists (n'_i)_{i=1,\dots S}\ {\rm s.  t.\
} \big( \gW_{L_1}(E, y_1, w, (n'_i))\ {\rm and\ }
\gH_{L_1}(E,y_1,m,(n'_i)) \ \big) \nonumber \\ \ \ {\rm or\ } \big(
\gW_{L_1}(E, y_2, w, (n'_i))\ {\rm and\ } \gH_{L_1}(E,y_2,m,(n'_i)) \
\big) \Big\} \geq 1-L_1^{-p}\ .\label{prob2}
\end{eqnarray}
\end{lem}

\begin{proof}
  Let $y_1$ and $y_2$ be such that $\pa y_1 - y_2 \pa_\infty > 2L_1$.
  We denote by $\overline{\gW}$ (resp. $\overline{\gH}$) the
  complementary events of $\gW$ (resp. $\gH$). We start by estimating
  the probability of the complement of the event that appears in
  \eqref{prob2}
  \begin{eqnarray}
    \lefteqn{ \bbP\Big\{\exists E\in I,\ \forall (n_i)_{i=1,\dots,
      S}\ \big( 
      \overline\gW_{L_1}(E,y_1,w,(n_i))\ {\rm or}\
      \overline\gH_{L_1}(E, 
      y_1,m,(n_i)) \big)} \nonumber \\
    & & {\rm and}\ \big(\overline\gW_{L_1}(E,y_2,w,(n_i))\
{\rm or}\ \overline\gH_{L_1}(E, y_2,m,(n_i))\big) \Big\} 
\nonumber \\
& \leq & \bbP\{ \exists E\in I,\ \forall (n_i) \
      \overline{\gW}_{L_1}(E, 
y_1, m, (n_i))\ {\rm and\ } \overline{\gW}_{L_1}(E,y_2, m, (n_i))\}
\nonumber \\
&  & +\bbP\{ \exists E\in I,\ \forall (n_i) \ \overline{\gH}_{L_1}(E,
y_1, w, (n_i))\ {\rm and\ } \overline{\gH}_{L_1}(E,y_2, m, (n_i))\}
\nonumber \\
&  & +\bbP\{ \exists E\in I,\ \forall (n_i) \ \overline{\gW}_{L_1}(E, 
y_1, w, (n_i))\ {\rm and\ } \overline{\gH}_{L_1}(E,y_2, m, (n_i))\}
\nonumber \\
&  & +\bbP\{ \exists E\in I,\ \forall (n_i) \ \overline{\gH}_{L_1}(E, 
y_1, m, (n_i))\ {\rm and\ } \overline{\gW}_{L_1}(E,y_2, w, (n_i))\}
\nonumber \\ 
& \leq  & \!\!\!\!\!\sum_{(n_i)\in{\cal V}}\!\!\!\bbP \Big\{\exists E\in I,
      \ \forall j,k\in\{1,\dots,N-S\}, \  
\nonumber \\
& &  \ \ \ \ \ \ \ \ 
\dist(\sigma(H_{\Lambda_{\frac{n_j L_1}{N}} (y_1)}),E)<L_1^{-w} {\rm
  and\ } \dist(\sigma(H_{\Lambda_{\frac{n_k L_1}{N}} 
  (y_2)}),E)<L_1^{-w} \! \Big\}   \nonumber \\
& & \ + 3\bbP\{\exists E\in I,\  \forall (n_i), \
      \overline{\gH}_{L_1}(E,y_1, m, (n_i))   
\} \ ,\label{prob3}
\end{eqnarray}
where ${\cal V}=\{(n_i)\, | \,1\leq n_1<n_2<\dots<n_{N-S}\leq N-1\}$.
The probability in the first term in \eqref{prob3} is bounded from
above by
\begin{eqnarray}
\bbP\{\exists E\in I,\ \Tr ({\bE}_1(J_w))( \Tr ({\bE}_2(J_w)) ) \geq 1
\}\ ,\label{jointwegner}
\end{eqnarray}
where $J_w=(E-L_1^{-w}, E+L_1^{-w})$ and $\bE_1$, $\bE_2$ are,
respectively, the spectral family of $H_{ \Lambda_{{n_1 L_1}/{N}}(y_1)
}$ and $H_{\Lambda_{{n_2 L_1}/{N}} (y_2)}$. Now \eqref{jointwegner} is
bounded from above by
\begin{eqnarray}
\lefteqn{ \bbP\left\{ \int_I {\Tr}({\bE}_1(J_w)){\Tr}(\bE_2(J_w))\,
dE\geq \frac{L_1^{-w}}{2}\right\} } \nonumber \\ 
& \leq & \int
d\bbP(\omega) \unc_{ \{\int_I \Tr ({\bE}_1(J_w)) \Tr ({\bE}_2(J_w))\,
dE \geq \frac{L_1^{-w}}{2} \} } \nonumber \\ 
&\leq & \int
d\bbP(\omega) 2L_1^{w} \int_I \Tr ({\bE}_1(J_w))\Tr ({\bE}_2(J_w))
\nonumber \\ & \leq & 2L_1^w \int_I \bbE \{ \Tr ({\bE}_1(J_w))\Tr
({\bE}_2(J_w)) \} dE \nonumber \\ 
& \leq & 2C_W\vert I \vert L_1^w
L_1^{-2w} \vert \Lambda_{\frac{n_1 L_1}{N}}(y_1) \vert\, \vert
\Lambda_{\frac{n_2 L_1}{N}}(y_2) \vert
\label{1term1} \\ 
& \leq & 2^{2d+1}C_W\vert I \vert L_1^{-w+2d} \label{1term2}\ . 
\end{eqnarray}
In inequality \eqref{1term1} we have used
Proposition~\ref{CHMestimate}.  The second term in \eqref{prob3} is
estimated as follows
\begin{eqnarray}
\lefteqn{
\bbP\{\exists E\in I,\ \forall (n_i),\
\overline{\gH}_{L_1}(E,y_1,m,(n_i)) \} 
} \nonumber  \\ 
& \leq & \bbP\big\{\exists E\in I, \exists 1\leq n_1<\dots<
n_{N-S}\leq 
N-1,\ \exists (z_i)_{i=1,\dots,N-S},
\nonumber \\
& &  \ \ \ \ 
\ z_i\in \gF_{L_1}(y_1,n_i),\ {\rm s.t.\ }
\Lambda_{L_0}(z_i) { \rm \ are\ not\ }(m,E){\rm 
-good\ boxes} \big\} \nonumber \\
& \leq & \begin{pmatrix} N-1 \\ N-S \end{pmatrix} ({\rm
Card}(\gF_{L_1}))^{N-S} \nonumber\\ & & \times \bbP\{\exists
E\in I,\ \forall 
i\in\{1,\dots,N-S\},\ \Lambda_{L_0}(z_i)
{\rm \ are\ not\ } (m,E) {\rm -good \ boxes}\} \ \nonumber \\
& \leq &
c(N,S,d)L_1^{(1-\frac{1}{\alpha})(d-1)(N-S)}L_0^{-p\theta(N-S,\alpha)} 
\ ,\label{2term1} 
\end{eqnarray}
for some uniform constant $c(N,S,d)$ depending only on $N$, $S$ and
$d$. In the last inequality we have used Assumption (A4).  Now,
\eqref{1term2} and \eqref{2term1} give
\begin{eqnarray*}
\lefteqn{\!\!\!\!\!\!\!\!\!\!\!\!\!\!\!\!\! \bbP\Bigg\{ \forall
E\in I,\ \exists (n'_i)_{i=1,\dots S}\ {\rm s. 
t.\ } \Big( \gW_{L_1}(E, y_1, w, (n'_i))\ {\rm and\ }
\gH_{L_1}(E,y_1,m,(n'_i)) \ \Big) } \\
& & \!\!\!\!\!\!\!\!\!\!\!\!\!\!\!\!\!\!\!\!\!\!\!\!{\rm or\ }  
\Big(\gW_{L_1}(E, y_2, w, (n'_i))\ {\rm and\ } 
\gH_{L_1}(E,y_2,m,(n'_i)) \ \Big) \Bigg\} \\
&\geq & 1-2^{2d+1}\begin{pmatrix} N-1 \\ N-S \end{pmatrix} C_W |I| 
L_1^{-w+2d} \\
& & -3c(N,S,d)N^{\frac{p\theta(N-S,\alpha)}{\alpha}}
L_1^{ (1-\frac{1}{\alpha})(d-1)(N-S) \nonumber-
\frac{p\theta(N-S,\alpha)}{\alpha} } \\
& \geq & 1-L_1^{-p}\ .
\end{eqnarray*}
\end{proof}

%%% Local Variables: 
%%% mode: latex
%%% TeX-master: "newdyn"
%%% End: 

\section*{Acknowledgments}
The authors want to thank J.M.\ Combes, P.D.\ Hislop, H.\ 
Leschke, R. Montcho and E.\ Mourre for helpful and stimulating
discussions about various aspects of this paper. 
In particular, T.\ Hupfer is acknowledged for pointing out an error in
a previous version.
JMB also thanks S.\ 
De Bievre and F.\ Germinet for interesting remarks about the
multi-scale analysis of von Dreifus and Klein. WF is grateful to J.M.\ 
Combes for hospitality and support at the Centre de Physique
Th\'eorique, CNRS, Marseille and at the Universit\'e de Toulon et du
Var, and JMB thanks H. Leschke for hospitality at the Universit\"at
Erlangen-N\"urnberg where part of this work was done. This work was
financially supported by the European Union through TMR network
FMRX-CT 96-0001 (JMB) and by the Deutsche Forschungsgemeinschaft (WF).

%%% Local Variables: 
%%% mode: latex
%%% TeX-master: "newdyn"
%%% End: 


\begin{thebibliography}{99} 
\bibitem{AG} M. Aizenman, G.M. Graf: Localization bounds for an
  electron gas, {\it J. Phys. A} {\bf 31}, 6783--6806 (1998).
\bibitem{AM} M. Aizenman, S. Molchanov: Localization at large disorder
  and at extreme energies: an elementary derivation, {\it Commun.
    Math. Phys.} {\bf 157}, 245--278 (1993).  
\bibitem{anderson} P.W. Anderson: Absence of diffusion in certain
  random lattices, {\it Phys. Rev.} {\bf 109}, 1492--1505 (1958).  
\bibitem{BCH} J.M. Barbaroux, J.M. Combes, P.D.
  Hislop: Localization near band edges for random Schr\"odinger
  operators, {\it Helv. Phys. Acta} {\bf 70}, 16--43 (1997).
\bibitem{BCM} J.M. Barbaroux,
  J.M. Combes, R. Montcho: Remarks on the relation between quantum
  dynamics and fractal spectra, {\it J. Math. Anal. Appl.} {\bf 213},
  698--722 (1997).  
\bibitem{BvES} J. Bellissard, A. Van Elst, H. Schulz-Baldes: The
  non-commutative geometry of the quantum Hall effect, {\it J. Math.
    Phys.} {\bf 35}, 5373--5451 (1994).  
\bibitem{CaLa90} R. Carmona, J. Lacroix:
  Spectral theory of random Schr\"odinger operators,
  Boston, Birkh\"auser, 1990.
\bibitem{CH1} J.M. Combes, P.D. Hislop: Localization for some
  continuous random Hamiltonians in $d$-dimensions, {\it J. Funct.
    Anal.} {\bf 124}, 149--180 (1994). 
\bibitem{CHM1} J.M.  Combes, P.D. Hislop, E. Mourre: Correlated Wegner
  inequalities for random Schr\"odinger operators, {\it Contemporary
    Mathematics} {\bf 217}, 191--203 (1998). 
\bibitem{dRJLS} R. Del Rio, S. Jitomirskaya, Y. Last, B. Simon:
  Operators with singular continuous spectrum IV: Hausdorff
  dimensions, rank one perturbations and localization, {\it J.
    d'Analyse Math.} {\bf 69}, 153--200 (1996).
\bibitem{vondreifusklein1} H. von Dreifus, A. Klein: A new proof of
  localization in the Anderson tight binding Model, {\it
  Commun. Math. Phys.} {\bf 124}, 285--299 (1989). 
\bibitem{vondreifusklein2} H. von Dreifus, A. Klein: Localization for
  random Schr\"odinger operators with correlated potentials, {\it
  Commun. Math. Phys.} {\bf 140}, 133--147 (1991).
\bibitem{FLM} W. Fischer, H. Leschke, P. M\"uller: Localization by
  Gaussian random potentials in multi-dimensional continuous space. 
  \emph{Preprint}, to be submitted to J.\ Stat.\ Phys.\ (1999).  
\bibitem{FS} J. Fr\"ohlich, T. Spencer: Absence of diffusion in the
  Anderson tight binding model for large disorder or low energy, {\it
  Commun. Math. Phys.} {\bf 88}, 151--184 (1983). 
\bibitem {debievregerminet} F. Germinet, S. De Bi\`evre: Dynamical
  localization for discrete and continuous random Schr\"odinger
  operators, {\it Commun. Math. Phys.} {\bf 194}, 323-341 (1998).
\bibitem{HS} P.D. Hislop, I.M. Sigal: Introduction to Spectral Theory
  with Applications to Schr\"odinger Operators, Applied Mathematical
  Sciences 113, New York, Springer, 1996.
\bibitem{holdenmartinelli} H. Holden, F. Martinelli: On absence of diffusion
 near the bottom of the spectrum for a random Schr\"odinger operator 
on $L^2(\bbR^\nu )$, {\it Commun. Math. Phys.} {\bf 93}, 197--217 (1984).
\bibitem{Kr2} W. Kirsch: Random
  Schr\"odinger Operators: a course, in Schr\"odinger Operators,
  S{\o}nderborg DK 1988, ed. H. Holden and A. Jensen. Lecture Notes in
  Physics Vol. 345, Berlin, Springer, 1989.  
\bibitem{KSS1} W. Kirsch, P. Stollman, G. Stolz: Anderson
  localization for Random Schr\"odinger operators with long range
  interactions, {\it Commun. Math. Phys.} {\bf 195}, 495--507 (1998). 
\bibitem{Kl1} F. Klopp:
  Localization for some continuous random Schr\"odinger operators,
  {\it Commun. Math. Phys. }  {\bf 167}, 553--569 (1995).
\bibitem{KS} H. Kunz, B. Souillard: Sur le spectre des op\'erateurs aux
  diff\'erences finies al\'eatoires, {\it Commun. Math. Phys.} {\bf
    78}, 201--246 (1980).
\bibitem{LiGr88} I.M. Lifshits, S.A. Gredeskul, L.A. Pastur:
  Introduction to the theory of disordered systems, New York, Wiley, 1988.
\bibitem{M} R. Montcho: Propri\'et\'es spectrales et dynamiques pour
  des op\'erateurs de Schr\"odinger \`a barri\`eres efficaces dans
  $L^2(\bbR^d)$, Thesis, Toulon, 1997. 
\bibitem{pa} L.A. Pastur: Spectra of random self-adjoint operators,
  {\it Russian Math. Surv.} {\bf 28}, 1--67 (1973).  
\bibitem{FP} L.A. 
  Pastur, A. Figotin: Spectra of random and almost-periodic
  operators, Berlin, Springer, 1992. 
\bibitem{radinsimon} C. Radin, B. Simon: Invariant domains for the
 time-dependent Schr\"odinger equation, {\it J. Diff. Equ.} {\bf 29},
 289--296 (1978).  
\bibitem{reedsimon3} M. Reed,
  B. Simon: Methods of Modern Mathematical Physics, III: Scattering
 Theory, New York, Academic, 1981.  
\bibitem{BS} H. Schulz-Baldes, J. Bellissard: Anomalous transport: a
  mathematical framework, {\it Rev. Math. Phys.} {\bf 10} 1--46 (1998).
\bibitem{ShEf84} B.I. Shklovskii, A.L. Efros:
  Electronic properties of doped semiconductors, Berlin, Springer, 1984.
\bibitem{thouless} D.J. Thouless: Electrons in disordered systems and
  the theory of localization, {\it Phys. Rep.} {\bf 13}, 93--142 (1974).


\end{thebibliography}
\end{document}